\documentstyle [12pt]{article}
\input psfig
\begin{document}

\newcommand{\bh}{\mbox{$\hat\beta$}}
\newcommand{\bhs}{\mbox{$\hat\beta^2$}}
\newcommand{\bhst}{\mbox{$\hat\beta^2/2$}}
\newcommand{\gh}{\mbox{$_3F_2$}}
\newcommand{\gio}{\mbox{$\gamma_{i_1}$}}
\newcommand{\git}{\mbox{$\gamma_{i_2}$}}
\newcommand{\de}{\mbox{$\delta$}}
\newcommand{\abot}{\mbox{{$\alpha\hat\beta \over 2$}}}
\newcommand{\B}{\mbox{{$\rm{B}$}}}
\newcommand{\K}{\mbox{{$\rm{K}$}}}
\newcommand{\wi}{{w_{i_1}}}
\newcommand{\wii}{{w_{i_2}}}
\newcommand{\wiii}{{w_{i_3}}}
\newcommand{\di}{{\delta_{i_1}}}
\newcommand{\dii}{{\delta_{i_2}}}
\newcommand{\diii}{{\delta_{i_3}}}
\newcommand{\nn}{\nonumber}
\newcommand{\ep}{\epsilon}
\newcommand{\eb}{\hat\beta^2}
\newcommand{\sk}{S_{K_{\pm}}(\hat\beta )}
\newcommand{\sj}{S_{J_{\pm}}(\hat\beta )}
\newcommand{\spl}{\bar{S}_{J_+}(\epsilon )}
\newcommand{\nsj}{\bar{S}_{J_{\pm}}(\hat\beta )}
\newcommand{\sm}{\bar{S}_{J_-}(\epsilon )}
\renewcommand{\theequation}{\thesection.\arabic{equation}}

\title{Short Distance Expansions \\ of Correlation Functions \\
in the Sine-Gordon Theory
}
\author{Robert M. Konik and Andr\'{e} LeClair \\
Newman Laboratory \\ Cornell
University \\ Ithaca, NY 14853}
\date{March 15, 1995}
\vskip 6pt
\maketitle
\thispagestyle{empty}

We examine the two-point correlation
functions of the fields exp(i$\alpha \Phi$) in the sine-Gordon theory
at all values of the coupling constant $\hat\beta$.
Using conformal perturbation theory, we
write down explicit integral expressions for every order of the
short distance expansion.
Using a novel technique analagous to dimensional
regularisation, we evaluate these integrals for the first few
orders finding expressions in terms of generalised hypergeometric
functions.  From these derived expressions, we examine the limiting forms
at the points where the sine-Gordon theory maps onto a
doubled
Ising and the Gross-Neveu SU(2) models.
In this way we recover the known
expansions of the spin and disorder fields about criticality in the Ising
model and the well known Kosterlitz-Thouless flows in the Gross-Neveu
SU(2) model.

\vspace{-7in}
\hspace{3.8in} CLNS 95/1326

\hspace{3.8in} hep-th/9503112

\setcounter{page}{1}
\pagebreak
\section {Introduction}

The most important generally unsolved problem in the subject of massive
integrable quantum field theory in two dimensions is the
computation of correlation functions.
A completely general approach to the study of the large distance
expansion is well known:  one can insert a multiparticle resolution
of the identity between fields and thus obtain an infinite integral
representation for the correlation function involving the form
factors.  Multiparticle form factors have been computed in a variety
of models by Smirnov \cite{smir}.  In principle, this form factor sum
completely characterizes the correlation function.  However the
complexity of the form factors makes this representation difficult
to utilize.

On the other hand, many quantum field theories have well defined
short distance expansions.  This short distance expansion is
essentially independent of the large distance expansion discussed above;
it is indeed extremely difficult to obtain one from the other.
In this paper we develop the short distance expansion of sine-Gordon
correlation functions.  This is also an infinite integral representation,
though the integrals can be much more compactly written than the large
distance expansion involving the form factors.  It is our aim to
demonstrate that this short distance expansion is tractable.

The form of the two point correlation functions of the fields
exp(i$\alpha \Phi$) in the sine-Gordon model at the free-fermion point
($SG_{ff}$) is well understood.  Because these fields in the $SG_{ff}$
can be mapped onto a doubling of the spin/disorder fields in the Ising
model (Ityzkson and Zuber \cite{iz}), the results  found
by Wu et al.\cite{wu}
allow the
expression of the correlators as solutions of a Painleve III non-linear
differential equation.
More recently, this result was extended in
\cite{ble}.
There, using a generalization of the techniques developed in
\cite{korepin},
it was shown
how
correlators of the fields exp(i$\alpha \Phi$), $0 < \alpha < 1$,
are characterized by
solutions to a non-linear equation, the sinh-Gordon equation.  The proof
of this result lay in expressing the large distance expansion of
the correlators as Fredholm determinants
through the use of form factors, and then deriving differential equations
for the Fredholm determinants.

In this paper we present new results characterizing correlators of
the above fields in
conformal perturbation theory for values of the sine-Gordon coupling,
$\hat \beta$, in the range $1 < \hat \beta < \sqrt2$.
(Here, $\hat{\beta}$ is related to the conventional sine-Gordon
coupling constant as  $\hat{\beta} = \beta/\sqrt{4\pi} $, the
free fermion point occuring at $\hat{\beta} = 1$.)
Because of the U(1) charge symmetry of the conformal limit
of the sine-Gordon theory, we focus upon
the following two correlators:
\begin{eqnarray}
G(\alpha , -\alpha ) & = & \langle \exp (i\alpha \Phi (z, \bar{z} ) )
\exp (-i\alpha \Phi (0));\\
G(\hat\beta /2 , \hat\beta /2) & = & \langle \exp (i\hat\beta\Phi (z,
\bar{z} )/2 )
\exp(i\hat\beta\Phi (0)/2) .
\end{eqnarray}
Expanding in powers of the mass parameter, $\lambda$,
we are able to write integral expressions
for every order of the short distance expansion.
However, not all orders are finite
for the given range of $\hat\beta$.
For the correlator
$G(\alpha, -\alpha)$,
IR singularities are absent
at $O(\lambda^{2n})$ given $\hat\beta^2 > 2 - 1/n$, while
for the correlator $G(\hat\beta /2,\hat\beta /2)$, IR singularities
do not appear at $O(\lambda^{2n+1})$ provided $\hat\beta^2 > (2n+1)/(n+1)$.
UV singularites are absent at every order for both correlators
provided $\hat\beta < \sqrt{2}$.

In principle it is possible to
evaluate the integrals at every order.  The singularities appear as poles
in the resulting functions: at lower
orders, gamma functions, at higher orders,
generalised hypergeometric functions.
However, the results at higher
orders quickly become intractable and not
particularly illuminating.  Thus we restrict ourselves to evaluating the
correlator, $G(\alpha ,-\alpha)$, to $O(\lambda^2)$, and the correlator,
$G(\hat\beta /2,\hat\beta /2)$, to $O(\lambda^3)$.
Indeed, $G(\hat\beta /2,\hat\beta /2)$ at $O(\lambda^3)$ is already
only finite for $\bhs > 3/2$, one-half of the full range of interest,
$1 < \bhs < 2$.

The integral expressions in the perturbative expansion do not arise from a
diagrammatic analysis.  Thus it might seem necessary to explicitly subtract
the contribution of the bubble diagrams.  (Without such a subtraction, the
terms in the expansion are formally divergent.)  However, the technique,
as first introduced by Dotsenko \cite{dot}, we use to evaluate the terms
renders them finite without explicit inclusion of the bubbles in a manner
analagous to
dimensional regularisation.  Rather than analytically continuing
the dimension of spacetime, we continue the parameter, $\hat\beta$.  In this
way not only
are the integrals rendered finite, the bubble diagrams evaluate to zero.

At particular points in the coupling, the sine-Gordon theory is
known to map onto familar theories.  At $\hat \beta = 1$, sine-Gordon maps
onto a doubled Ising
model, and at $\hat \beta = \sqrt2$, it
maps onto the SU(2) Gross-Neveu model.  We
examine our perturbative expansion at these points.
At both of these points
the perturbative series become
infinite and renormalisation is required.
At $\hat\beta = 1$
we introduce a novel renormalisation scheme to remove the
IR divergences.
At $\hat\beta = \sqrt{2}$ we employ a more conventional scheme
to take care of the UV divergences.
Having then performed the necessary
renormalisation, we obtain the known expansions of correlators for
the order/disorder fields in the Ising model as well as the known
Kosterlitz-Thouless flows in the Gross-Neveu SU(2) model.  This provides
some degree of confidence that our derived expressions away from the free
fermion point are correct.

\section{Overview of the Perturbative Expansion}
\setcounter{equation}{0}
The action for the sine-Gordon theory in Euclidean space-time
(our conventions are $z = {(t + ix)/2}, \bar z = {(t - ix)/2}$ and
$d^2z = -dtdx/2)$ is
\begin{equation}
\label{act}
S = S_{CFT} + S_{Pert.} =
-{1 \over 4 \pi} \int d^2 z \Bigl(\partial_z \Phi \partial_{\bar z} \Phi + 4
\lambda :cos(\hat \beta \Phi): )
\end{equation}
where, again, the parameter $\hat \beta$ is related to the conventionally
normalized coupling by $\hat \beta = {\beta \over \sqrt{4\pi}}$.  As shown
by Coleman \cite{cole}, the UV divergences in the theory for values of
$\hat \beta < \sqrt2$
(i.e. for values of $\hat \beta$ where the coupling $\lambda$ is
not irrelavent), may be removed by the straight-forward normal ordering
of the $\cos (\hat{\beta} \Phi  )$.
In the process, the coupling constant $\lambda$ is multiplicatively
renormalised.  As no wavefunction renormalisation is necessary, the scaling
dimensions of the fields remain unchanged by the introduction of the cosine
perturbation -
the dimensions are the same as those in the deep ultraviolet.
Because the structure of the field theory
is the same as its conformal limit,
it is meaningful to perturb about the free theory - indeed, we may use the
same labels for the fields in both the free and massive theories.
For a general discussion of conformal perturbation theory see \cite{cpt}.

The coupling constant, $\lambda$, in \ref{act} can be directly related to
the mass, $m$, of the asymptotic states, the solitons.  Zamolodchikov
\cite{Zam}, using results derived with
thermodynamic Bethe ansatz, demonstrated that
\begin{equation} \label{lamm}
\lambda = -\tilde{\lambda} ( \bh ) m^{2-\eb}
\end{equation}
where \footnote{Some caution must be used in applying Zamolodchikov's
result directly as he uses different conventions for both the action and
the propagator.}
\begin{equation}
\tilde{\lambda} (\bh ) = {\Gamma (\bhs/2 ) \over \Gamma (1-\bhs/2)}
\left[ {\sqrt{\pi} \Gamma (1/(2-\bhs ) \over \Gamma (\bhs /(4-2\bhs )}
\right]^{2-\eb}.
\end{equation}
We note that $\lambda = -m$ at $\bh = 1$.  This relation (\ref{lamm})
will prove to be
useful in our examination of the Ising model.

Consider the correlation functions
\begin{equation}
G (\alpha,\alpha ') = \langle 0|:
\exp(i \alpha \Phi (z,\bar z))::\exp(i\alpha ' \Phi (0)):|0\rangle .
\end{equation}
These correlation functions are highly non-trivial.
Even at the free-fermion point, $\hat\beta = 1$, the bosonisation
relations,
$:\exp({\pm} \Phi): = \Psi_{\pm}
\Psi_{\mp}$, where the $\Psi$'s are the spinor components in the
massive Thirring model, imply the fields, $\exp(i \alpha \Phi)$,
are not simply representable as fermions.

The perturbative expansion of these correlators in a path integral
formulation is \small
\begin{eqnarray}
G (\alpha,\alpha ') & = &  {\int D\Phi
e^{-S(\Phi)}e^{i\alpha\Phi(z)}e^{i\alpha'\Phi(0)} \over {\int
D\Phi e^{-S(\Phi)}}} \\
\nonumber && \hspace{-.6in} = ~{\int D\Phi
e^{-S_{CFT}(\Phi)}e^{{\lambda \over \pi} \int d^2w~cos(\hat \beta \Phi(w))}
e^{i\alpha\Phi(z)}e^{i\alpha'\Phi(0)} \over {\int
D\Phi e^{-S_{CFT}(\Phi)}
e^{{\lambda \over \pi} \int d^2w~cos(\hat \beta \Phi(w))}}} \\
\nonumber && \hspace{-.6in} = ~{\displaystyle{\sum_{n=0}^{\infty}}
{1 \over n!}
{\left(\lambda \over \pi \right)}^n \int d^2w_i
< \! cos(\hat \beta \Phi(w_1))
\cdots cos(\hat \beta \Phi(w_n))
e^{i\alpha\Phi(z)}e^{i\alpha'\Phi(0)} \! >_{CFT} \over
\displaystyle{\sum_{n=0}^{\infty}} {1 \over n!}
{\left(\lambda \over \pi \right)}^n \int d^2w_i
< \! cos(\hat \beta \Phi(w_1))
\cdots cos(\hat \beta \Phi(w_n)) \! >_{CFT}} \\
\nonumber && \hspace{-.6in} =~ {\displaystyle{\sum_{n=0}^{\infty}}
{1 \over n!}
{\left(\lambda \over 2\pi \right)}^n \int d^2w_i
\displaystyle{\sum_{l_k = \pm 1}} <  \! e^{i l_1 \hat \beta \Phi(w_1)}
\cdots e^{i l_n \hat \beta \Phi(w_n)}
e^{i\alpha\Phi(z)}e^{i\alpha'\Phi(0)} \! >_{CFT} \over
\displaystyle{\sum_{n=0}^{\infty}}
{1 \over n!} {\left(\lambda \over 2\pi \right)}^n \int d^2w_i
\displaystyle{\sum_{l_k = \pm 1}} < \! e^{i l_1 \hat \beta \Phi(w_1)}
\cdots e^{i l_n \hat \beta \Phi(w_n)} \! >_{CFT}}
\end{eqnarray}
\normalsize
where we have expanded $e^{-S_{Pert}}$ in powers of $\lambda$ and
$<>_{CFT}$ indicates the correlator is to be evaluated in the conformal
limit.  The advantage of expressing $G (\alpha,\alpha ')$ in this form is
that the correlators of vertex operators in the conformal limit,
i.e. $<>_{CFT}$, are easily evaluated.
The denominators of these expressions,
representing the bubble diagrams in the theory,
will prove to be important in proving the convergence of the terms in the
perturbative expansion.

To evaluate the resulting conformal correlators we need to specify the
free boson propagator.  With the action as given in \ref{act} the
propagator is
\begin{equation}
 \langle \Phi(z,\bar z)\Phi(0) \rangle = - \log (c z \bar z ) ,
\end{equation}
where $c$ is some arbitrary constant.  As our convention we set
$c = 1$.  With this propagator the OPE of two vertex operators
$:e^{i\alpha\Phi(z,\bar z)}::e^{i\beta\Phi(0)}:$, is then equal to
\begin{equation}
:e^{i\alpha\Phi(z,\bar z)}::e^{i\beta\Phi(0)}: =
\left( |z|^2 \right)^{\alpha\beta}
:e^{i(\alpha + \beta)\Phi(z,\bar z)}: + \cdots.
\end{equation}
Using this OPE, the terms of the form $< \! e^{i\alpha_1\Phi(z_1)}\cdots
e^{i\alpha_n\Phi(z_n)} \! >_{CFT}$ are then given by
\begin{equation} \label{confcor}
< \! e^{i\alpha_1\Phi(z_1)} \cdots e^{i\alpha_n\Phi(z_n)} \! >_{CFT} =
\delta (\sum \alpha_i)
\prod_{i<j} \left( |z_i - z_j|^2 \right)^{\alpha_i \alpha_j}.
\end{equation}
That $<>_{CFT}$ vanishes unless $\sum \alpha_i
= 0$ is as consequence of the U(1) symmetry of the
 free massless theory.

Given this U(1) symmetry, only certain values of $\alpha$ and $\alpha '$
lead to non-vanishing correlators, $G (\alpha,\alpha')$, in the
massive theory in perturbation theory.
We examine two in particular, $G (\alpha,-\alpha)$
and $G (\hat\beta /2,\hat\beta /2)$.  The first has non-vanishing
terms at all even orders in the perturbation, the second at all odd
orders.  One reason for choosing to examine these correlators in particular
is their relation to the Ising spin and disorder field
correlators, $< \! \sigma(z,\bar z) \sigma(0) \! >$ and $< \! \mu(z,\bar
z ) \mu(0) \!>$,
respectively.  At $\hat \beta = 1$, these correlators are related to the
$G$'s via
\begin{equation}
< \! \sigma(z,\bar z) \sigma(0) \! >^2 =
< \! \sin({\Phi(z,\bar z) \over 2}) \sin({\Phi(0) \over 2}) \! > =
{1 \over 2} \left( G ({1 \over 2},-{1 \over 2}) -
G ({1 \over 2},{1 \over 2}) \right)
\end{equation}
and
\begin{equation}
< \! \mu(z,\bar z) \mu(0) \! >^2 =
< \! \cos({\Phi(z,\bar z) \over 2}) \cos({\Phi(0) \over 2}) \! > =
{1 \over 2} \left( G ({1 \over 2},-{1 \over 2}) +
G ({1 \over 2},{1 \over 2}) \right) ,
\end{equation}
where we have used $G (\alpha,\alpha ') = G (-\alpha,-\alpha ')$.

The perturbative expansion for $G (\alpha,-\alpha)$ and
$G (\hat\beta /2,\hat\beta /2)$, using \ref{confcor} and the U(1) symmetry,
simplifies to
\begin{equation}
G (\alpha , -\alpha) =
{\displaystyle{\sum_{n=0}^{\infty}}
{1 \over n!}{\left(\lambda \over 2 \pi \right)}^{2n}
{2n \choose n}
\int d^2w_1 \cdots d^2w_{2n}
\omega (\hat\beta,w_i) \tau (\alpha,\hat\beta,w_i) \over
\displaystyle{\sum_{n=0}^{\infty}} {1 \over n!}
{\left(\lambda \over 2\pi \right)}^{2n}
{2n \choose n}
\int d^2w_1 \cdots d^2w_{2n} \omega (\hat\beta,w_i)}
\end{equation}

\noindent and

\begin{equation}
G ({\hat \beta \over 2},{\hat \beta \over 2}) =
{\displaystyle{\sum_{n=0}^{\infty}}
{1 \over n!}
{\left(\lambda \over 2 \pi \right)}^{2n+1} \! \!
\int d^2w_1 \cdots d^2w_{2n+1}
{2n+1 \choose n+1}
\chi (\hat\beta,w_i) \psi (\hat\beta,w_i) \over
\displaystyle{\sum_{n=0}^{\infty}} {1 \over n!}
{\left(\lambda \over 2\pi \right)}^{2n}
{2n \choose n}
\int d^2w_1 \cdots d^2w_{2n} \omega (\hat\beta,w_i)}
\end{equation}

\noindent
where the functions $\omega$, $\tau$, $\chi$, and $\psi$ are given by
\begin{eqnarray}
\omega(\hat\beta,w_i)  \!\!\! & = & \!\!\!
\prod_{1 \leq i \not= j \leq n \atop 2n \geq i \ne j > n}
(|w_i - w_j|^2)^{\hat\beta^2}
\prod_{1 \leq i \leq n < j \leq 2n} (|w_i - w_j|^2)^{-\hat \beta^2}, \\
\nonumber \tau(\alpha,\hat\beta,w_i) & = & |z^{-2}|^{\alpha^2}
\prod_{1 \leq i \leq n}(|w_i - z|^2|w_i|^{-2})^{\alpha\hat\beta}
\prod_{n < i \leq 2n}(|w_i - z|^{-2} |w_i|^2)^{\alpha\hat\beta}, \\
& &\\
\chi(\hat\beta,w_i) \!\!\! & = & \!\!\!
\prod_{1 \leq i \not= j \leq n \atop {2n+1} \geq i \ne j > n}
(|w_i - w_j|^2)^{\hat\beta^2}
\prod_{1 \leq i \leq n < j \leq 2n+1} (|w_i - w_j|^2)^{-\hat \beta^2}, \\
\nonumber \psi(\hat\beta,w_i) \!\!\! & = & \!\!\!
|z^2|^{\hat\beta \over 4}
\prod_{1 \leq i \leq n}(|w_i - z|^2|w_i|^{2})^{\hat\beta^2 \over 2}
\prod_{n < i \leq 2n+1}(|w_i - z|^{2} |w_i|^{2})^{-{\hat\beta^2 \over
2}}.\\
\end{eqnarray}

\def\betah{\hat{\beta}}

\section{Convergence of Terms in Perturbative Expansion}
\setcounter{equation}{0}
In this section we examine the singularities that appear in the
perturbative expansions
of the correlators $G (\alpha, -\alpha)$ and $G (\hat{\beta} /2,
\hat{\beta}/2 )$.
We find that $G(\alpha, -\alpha)$
at $O(\lambda^{2n})$ is UV finite for $\bhs <2$ and IR finite for
$\bhs > 2 - 1/n$, and that $G(\hat{\beta}/2 , \hat{\beta}/2)$
at $O(\lambda^{2n+1})$ is again UV finite for $\bhs <2 $ but IR finite
for $\bhs > (2n+1)/(n+1)$.
We give a proof of this beginning with $G (\alpha , - \alpha)$.

To show UV finiteness, we consider the contribution to
$G (\alpha , -\alpha)$ at $O(\lambda^{2n})$.  It takes the form
\begin{eqnarray} \label{term}
&& \nn \int d^2w_1 \cdots d^2w_{2n} \langle e^{i\alpha \Phi(z)}
e^{-i\alpha \Phi(0)} e^{i\betah \Phi (w_1)} e^{-i\betah \Phi (w_2)} \cdots
e^{i\betah \Phi (w_{2n-1})} e^{-i\betah \Phi (w_{2n})} \rangle \\
&&\hspace{1in} -~\hbox{\rm disconnected pieces} .
\end{eqnarray}
The leading order UV singularity occurs as $w_1 \rightarrow w_2$,
$w_3 \rightarrow w_4$, $\ldots$, $w_{2n-1} \rightarrow w_{2n}$.  These
singularities are governed by the operator product expansion
\begin{eqnarray} \label{ope1}
:e^{i\betah \Phi(x)}::e^{-i\betah \Phi(y)}: & = & |x - y|^{-2\hat\beta^2}
:e^{i\betah(\Phi(x) - \Phi(y))} : + ~\hbox{\rm finite pieces} \\
\nn && \hspace{-.5in} =~ |x-y|^{-2\hat\beta^2}
\left(I - \hat\beta^2 |x-y|^2 :\partial_{y}\Phi (y)
\partial_{\bar y}\Phi (y) : + \cdots \right) ,
\end{eqnarray}
where in the second line pieces such as $(x-y) \partial_y\Phi(y)$ and its
antiholomorphic counterpart are discarded.  Such pieces, when the OPE is
inserted into the correlator, integrate out.

The leading order singularity in this term
does not come from the contribution
of the identity operator in the above OPE; rather this term leads
to a disconnected piece which is subsequently subtracted off.  Thus the
leading order singularity that contributes is $|x-y|^{2-2\hat\beta^2}$.
So the
leading UV singularity at $O(\lambda^{2n})$ in $G (\alpha , -\alpha)$
has dimension
\begin{equation}
d_{UV} = n (2-2\bhs + 2) .
\end{equation}
The $n$ arises as there n approaches $w_i \rightarrow w_{i+1}$, the extra 2
from the integration $\int d^2(w_i - w_{i+1})$.  Hence for $\bhs <2$, i.e.
$d_{UV} > 0 $, the terms are UV finite,
as is expected from Coleman \cite{cole}.

The reader may worry that stronger divergences are introduced considering
the approaches $w_1 \rightarrow w_3$, $w_5 \rightarrow w_7$, and so
forth in addition to
$w_1 \rightarrow w_2$, $w_3 \rightarrow w_4$, etc.
But such additional
singularities are governed by the OPE
\begin{eqnarray} \label{ope2}
&& :\partial_x\Phi(x) \partial_{\bar x}\Phi(x):
:\partial_y\Phi(y) \partial_{\bar y}\Phi(y): = \\
\nn && \hspace{.5in} |x-y|^{-4}\left( I + |x-y|^2 :\partial_y\Phi(y)
\partial_{\bar y}\Phi(y) + \cdots \right) .
\end{eqnarray}
Again the leading term in the OPE is disconnected.  So its leading singular
contribution is $|x-y|^{-2}$.
But this is exactly cancelled by the integration
$\int d^2(x-y)$ and so the overall UV singularity becomes no greater.

When all the $w_i$'s are allowed to approach one another, successively
substituting the above OPEs into the correlator in \ref{term} leads
one eventually to the correlator
\begin{equation} \nn
\langle e^{i\alpha\Phi(z)} e^{-\alpha\Phi(0)}
:\partial_w\Phi(w) \partial_{\bar w}\Phi(\bar w ): \rangle .
\end{equation}
But this correlator, as can be easily checked, introduces no additional
UV singularities as $w$ approaches $z$ or 0.  Thus
allowing in addition the $w_i$'s to approach $z$ or 0 does not change the
UV properties of the perturbative expansion.

To demonstrate that $G (\alpha , -\alpha)$ is IR finite at
$O(\lambda^{2n})$ for $\bhs > 2 - 1/n$, we recast each term in the
perturbative expansion by making the conformal transformation
\begin{equation}
\nn w_i \rightarrow 1/w_i .
\end{equation}
The perturbative terms then appear as
\begin{eqnarray}
&& \nn \int |w_1|^{2\hat\beta^2 - 4} d^2w_1 \cdots
|w_{2n}|^{2\hat\beta^2 -4} d^2w_{2n}
\times \\
&& \nn\hspace{.5in}\langle e^{i\alpha \Phi(z)}
e^{-i\alpha \Phi(0)} e^{i\betah \Phi (w_1)} e^{-i\betah \Phi (w_2)} \cdots
e^{i\betah \Phi (w_{2n-1})} e^{-i\betah \Phi (w_{2n})} \rangle \\
&&\hspace{1.5in} -~\hbox{\rm disconnected pieces} .
\end{eqnarray}
The IR behaviour of this integral is determined by sending $w_i \rightarrow
0$ for all i.  The singular behaviour of the correlator
in the above integral
can be extracted by first sending $w_1 \rightarrow w_2$,
$w_3 \rightarrow w_4$, $\ldots$, $w_{2n-1} \rightarrow w_{2n}$, then
$w_1 \rightarrow w_3$, $w_5 \rightarrow w_7$, $\ldots$, $w_{2n-3}
\rightarrow w_{2n-1}$, and so on.  Using the OPEs given in \ref{ope1}
and \ref{ope2} and discarding, as before, the disconnected pieces associated
with the identity, the dimension of the leading IR singularity is
\begin{equation}
d_{IR} = \left[ 2n \left(2\bhs -2\right)\right] +
\left[2 - 2n\bhs \right] .
\end{equation}
The latter term in $d_{IR}$ is the contribution coming from the correlator,
the former term from $\int |w_i|^{2\hat\beta^2 - 4} d^2w_i$.
It is important to not include the singularities that arise from the OPEs
involving $e^{i\alpha\Phi(0)}$.  These singularities are properly UV.
Then for the term to be
IR finite, we need $d_{IR} > 0$.  Thus
$\bhs$ must be greater than $2 - 1/n$.

To check the UV finiteness of $G (\betah /2 , \betah /2)$, a similiar
argument is used to the one above.  At $O(\lambda^{2n+1})$ the term in the
perturbative expansion appears as
\begin{eqnarray}
&& \nn \int d^2w_1 \cdots d^2w_{2n+1} \times \\
\nn && \hspace{.3in} \langle e^{i\betah /2 \Phi(z)}
e^{i\betah /2 \Phi(0)} e^{-i\betah \Phi (w_1)} e^{i\betah\Phi (w_2)}
e^{-i\betah \Phi (w_3)} \cdots
e^{i\betah \Phi (w_{2n})} e^{-i\betah \Phi (w_{2n+1})} \rangle \\
&&\hspace{1in} -~\hbox{\rm disconnected pieces} .
\end{eqnarray}
The most singular pieces of this expression come when $w_2 \rightarrow w_3$,
$w_4 \rightarrow w_5$, $\ldots$, $w_{2n} \rightarrow w_{2n+1}$.  These
approaches are governed by the OPE in \ref{ope1}.  Thus the overall
dimension of the leading UV singularity is
\begin{equation}
d_{UV} = n(4-2\bhs ) ,
\end{equation}
and so again for $\bhs < 2$ the terms are UV finite.  Additional approaches,
say $w_2 \rightarrow w_4$, $w_6 \rightarrow w_8$, etc., are govered by the
OPE \ref{ope2} and so, as before, do not change $d_{UV}$.  If all the
$w_i$'s, $i>1$, are allowed to approach one another, say at $w$,
the succession of OPEs
leaves one to evaluate the correlator
\begin{equation} \label {ope} \nn
\langle e^{i\hat\beta/2\Phi(z)} e^{i\hat\beta/2\Phi(0)} e^{-i\beta\Phi(w_1)}
:\partial_w\Phi(w) \partial_{\bar w}\Phi(w): \rangle .
\end{equation}
But as $w \rightarrow w_1$, $0$, or $z$, no new UV singularities appear.

To obtain the IR behaviour of $G (\betah /2 , \betah /2)$ for $n>0$,
we begin by treating the term at $O(\lambda^{2n+1})$
like the term at $O(\lambda^{2n})$ in $G (\alpha , -\alpha)$,
contracting the last 2n vertex operators $e^{\pm i \betah \Phi(w)}$ in the
correlator.  However there is an additional contribution of
$2\bhs - 4$ to $d_{IR}$ in
this case that comes from letting $w \rightarrow w_1$ in \ref{ope}.
Thus the term is IR finite
if $\bhs > (2n+1)/(n+1)$.  For $n=0$ it is easy to see
that this formula still holds: the term at $O(\lambda )$ is IR finite
if $\bhs > 1$.

Though we have identified the ranges of $\hat\beta$ where the perturbative
expansion is divergent, we point out that the expressions we derive in the
next section will only be divergent for individual values of $\hat\beta$:
in general the expressions we will derive will be meromorphic functions
of $\hat\beta$.  However it is not clear how to interpret the analytically
continued expressions in regions of $\hat\beta$ where the original
integral expressions are divergent.  It is possible that
that this continuation does not capture all the relavent physics.
For example at $\bhs < 1$, it may be
the continued correlators do not reflect
the presence of quantum breathers.

\section{Evaluation of Terms of Perturbative Series}
\setcounter{equation}{0}
We begin by focusing on the integrals in the numerator of the
perturbative expansion (i.e. we ignore the contribution of the bubble
diagrams).  Although these integrals are formally divergent, the
techniques (as first developed by Dotsenko \cite{dot}) we
employ to evaluate them
render them finite in a manner somewhat analogous to dimensional
regularisation.  This handling of the infinities suggests the technique
takes into account the bubble diagrams.  This is born out by the
fact that when the
contribution of the bubbles are calculated explicitly using the same
techniques, they turn out to be identically zero.

The integrals we thus consider are of the form
\begin{equation}
I_{2n} = \int d^2w_1 \cdots d^2w_{2n}
\omega(\hat\beta,w_1,\cdots,w_{2n})
\tau(\alpha,\hat\beta,w_1,\cdots,w_{2n}),
\end{equation}
and
\begin{equation}
I_{2n+1} = \int d^2w_1 \cdots d^2w_{2n+1}
\chi(\hat\beta,w_1,\cdots,w_{2n+1})
\psi(\hat\beta,w_1,\cdots,w_{2n+1}).
\end{equation}
\noindent These integrals are, in their full generality, unwieldly.
Thus, in the end,
we only explicitly evaluate $I_1$, $I_2$, and $I_3$.  Indeed,
$I_3$ is already only finite for $\bhs >3/2$, a part of the
full range of interest.  However, we
will carry out the calculation for general n as far as seems reasonable.
We begin with $I_{2n}$.

\subsection{Evaluation of $I_{2n}$}

To make sense of $I_{2n}$ we must first abandon light cone coordinates,
writing it in the usual x-t
coordinates.  Doing so, and at the same time
scaling out the z-dependence, we obtain
\begin{eqnarray}
I_{2n} & = &
(-2)^{2n} |z|^{4n-2\alpha^2 + 2\hat\beta^2 n}
\int \! dt_1 dx_1 \cdots dt_{2n} dx_{2n}
\times \\
\nonumber & & \prod_{1 \leq i \neq j \leq n \atop n < i \neq j \leq 2n}
\!\!\!\!\!\! \left( (t_i - t_j)^2 + (x_i - x_j)^2 \right)^{\hat\beta}
\prod_{1 \leq i \leq n < j \leq 2n} \! \!
\left( (t_i - t_j)^2 + (x_i - x_j)^2 \right)^{-\hat\beta} \times \\
\nonumber & & \hspace{.1in}
\prod_{1 \leq i \leq n} \left( \left( (t_i - 1)^2 + x_i^2 \right)^2
\left( t_i^2 + x_i^2 \right)^{-2} \right)^{\alpha\hat\beta} \times \\
\nonumber & & \hspace{.08in} \prod_{n < i \leq 2n}
\left( \left( (t_{i} - 1)^2 + x_{i}^2 \right)^{-2}
\left( t_{i}^2 + x_{i}^2 \right)^2 \right)^{\alpha\hat\beta}.
\end{eqnarray}
\noindent Now the branch points in $I_{2n}$'s integrand
of the $x_i$'s lie on the imaginary
axis.  Thus the deformation of the x-contours pictured in Figure 1 is
permissible.
\begin{figure}
\par
\centerline{\psfig{figure=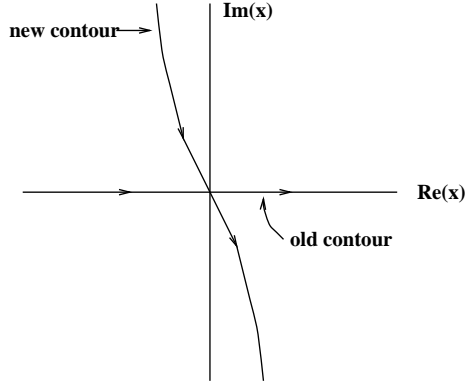,height=2in}}
\caption{A sketch of the deformation of the $x_i$ contours.}
\par
\end{figure}
This deformation corresponds to the following change of variables $x_i
\rightarrow -i e^{-i 2 \epsilon} x_i$ where $\epsilon$ is some small
positive parameter.  Given that the terms in the integral now take the
form $(t_i - t_j)^2 - e^{-i 4 \epsilon}(x_i - x_j)^2$, the following
changes of variables
$$ w_i^{\pm} = t_i \pm x_i$$
factorises these expressions leaving $I_{2n}$ in the form
\begin{eqnarray}
\lefteqn{I_{2n} =
(-2)^{2n} \left( i \over 2 \right)^{2n}
|z|^{4n-2\alpha^2 + 2\hat\beta^2 n} \times} \\
& & \nonumber \int \! dw^+_1 \cdots dw^+_{2n} dw^-_1
\cdots dw^-_{2n}
\theta(w^+_1,\cdots,w^+_{2n},\epsilon)
\theta(w^-_1,\cdots,w^-_{2n},-\epsilon)
\end{eqnarray}
\noindent where $\theta(w_1,\cdots,w_{2n},\epsilon)$ is given by
\begin{eqnarray}
\theta(w_1,\cdots,w_{2n},\epsilon) & = &\hspace{.1in} \left[ \prod_{i=1}^n
\left( w_i - 1 - i\epsilon\Delta_i \right)^{\alpha\hat\beta}
\left( w_i - i\epsilon\Delta_i \right)^{-\alpha\hat\beta} \right. \times \\
\nonumber & & \hspace{-1.5in} \left. \prod_{j=i+1}^n
\left( w_i - w_j - i\epsilon (\Delta_i - \Delta_j)
\right)^{\hat{\beta}^2}
\prod_{j=n+1}^{2n}
\left( w_i - w_j - i\epsilon (\Delta_i - \Delta_j)
\right)^{-\hat\beta^2} \right] \times \\
\nonumber & & \hspace{-1.5in} \left[ \prod_{i=n+1}^{2n}
\left( w_i - 1 - i\epsilon\Delta_i \right)^{-\alpha\hat\beta}
\left( w_i - i\epsilon\Delta_i \right)^{\alpha\hat\beta}
\prod_{j=i+1}^{2n}
\left( w_i - w_j - i\epsilon (\Delta_i - \Delta_j)
\right)^{\hat\beta^2} \right],
\end{eqnarray}
\noindent and $\Delta_i = w^+_i - w^-_i$.

\begin{figure}
\par
\centerline{\psfig{figure=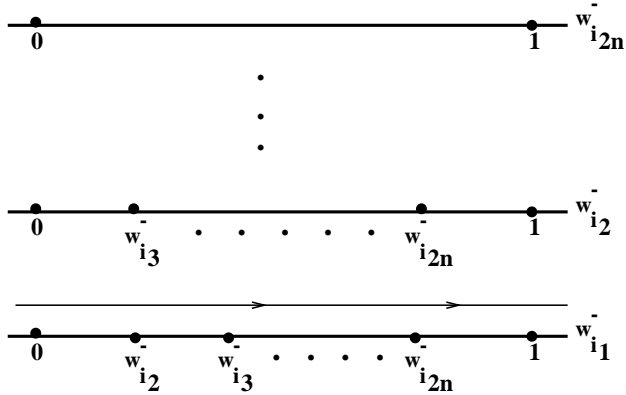,height=2in}}
\caption{The marked points on each of the contours represent the
branch points.  We have not drawn the contours for $w^-_{i_2}, \cdots ,
w^-_{i_{2n}}$ as they are irrelavent to the argument being made.}
\par
\end{figure}

Having made this set of changes of variable, we are faced with a set of
integrals that are represented by contours deformed around the branch
points of algebraic functions.  The signs of the coefficients of
$\epsilon$ in the integrand tell us how to make the deformations.
Focusing on the $w_i^-$-contours, the coefficients of the relavent
$\epsilon$'s are functions of the variables $w_i^+$.  For various values
of the $w_i^+$'s, the $w_i^-$-contours will enclose no poles, and the
corresponding contribution to $I_{2n}$ is 0.  To find these values we
consider a set {$w^+_i$} ordered as follows:
$$ w_{i_1}^+ > w_{i_2}^+ > \cdots > w_{i_{2n}}^+ .$$
We will show that if $w_{i_1}^+ > 1$ or $w_{i_{2n}}^+ < 0$, the
contribution to $I_{2n}$ is zero.

First suppose $w_{i_1}^+ > 1$.  Then if we perform the $w^-_i$ integrations
in ascending order of i, the set of contours for {$w_i^-$} appear as in
Figure 2.  Because the $w^-_{i_1}$-contour can be closed at $\infty$, the
contribution to $I_{2n}$ is 0.  Only the branch points of $w^-_i$ arising
from the terms of the form
$(w_i^- - a + i\epsilon f(w^+_k,w^-_k) )^\gamma$
are shown.  The branch points of $w_i^-$ arising from terms of the form
$(w_i^+ - a + i\epsilon f(w^+_k,w^-_k) )^\gamma$,
i.e. terms where $w_i^-$ appears multiplied by $\epsilon$,
are along the imaginary axis of $w_i^-$ and
are taken to $\pm i\infty$ as $\epsilon \rightarrow 0$.  Thus they do not
affect the deformation of the contour.

Now suppose $w^+_{i_n} < 0$.  The set of contours for {$w^+_i$} then
appear as in Figure 3.
\begin{figure}
\par
\centerline{\psfig{figure=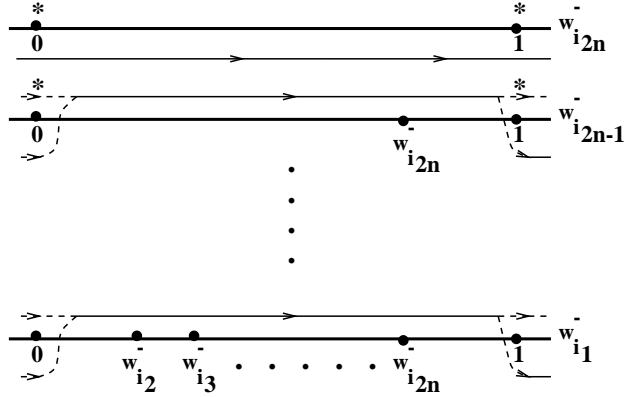,height=2in}}
\caption{Pictured are the set of contours for {$w_i^-$} when $w^+_{2n} < 0$.
The dashed lines indicate the contours may follow either path, depending
on the specific values of the $w_i^-$'s.  The $*$'s mark branch points
arising from prior integrations, as explained in the text.}
\par
\end{figure}
Because the $w^-_{2n}$ contour can be closed in the lower half plane, the
contribution to $I_{2n}$ is again 0.

In addition to the branch points that arise directly from the terms
$(w^-_i - a + i\epsilon f(w^+_k,w^-_k))$, branch points will also appear
because of prior integrations.  For example, after performing the first
2n-1 $w^-$-integrations, $I_{2n}$ is left in the form
\begin{eqnarray}
I_{2n} & = &\int dw^+_1 \cdots dw^+_{2n}dw^-_{2n}
\theta(w^+_1,\cdots,w^+_{2n},\epsilon) \times \\
& & \nonumber
\left( w^-_{2n} - 1 + i\epsilon\Delta_{2n} \right)^{-\alpha\hat\beta}
\left( w^-_{2n} + i\epsilon\Delta_{2n} \right)^{\alpha\hat\beta}
g(w^-_{2n} - i\epsilon ),
\end{eqnarray}
where g is a generalised hypergeometric function with branch points at 0
and 1.  These branch points arise from integrating terms of the form
$\left( w^-_k - w^-_{2n} + i\epsilon(\Delta_k - \Delta_{i_{2n}})
\right)^\gamma$, $k \neq i_{2n}$.  As the coefficent of $\epsilon$ in
this term is positive when $w^-_k = w^-_{i_{2n}}$ (as
$w^+_k - w^+_{i_{2n}} > 0$), g depends on $w^-_{2n} - i\epsilon$ as
indicated.  Hence the $w_{i_{2n}}$-contour
flows underneath the branch points
of g.  The importance of all of this is, of course, that g's branch
points do not interfere with the deformation of the contour that takes
the contribution to 0.

Thus is has been shown that only for values of {$w^+_i$} constrained to
lie between 0 and 1 is a non-zero contribution to $I_{2n}$ made.  As the
are n! ways of ordering the $w_i^+$'s, there are n! cases to consider,
each having the general form
$$1 > w^+_{i_1} > \cdots > w^+_{i_{2n}} > 0 .$$
The contours for $w^-_i$'s in this general case appear as in Figure 4.
\begin{figure}
\par
\centerline{\psfig{figure=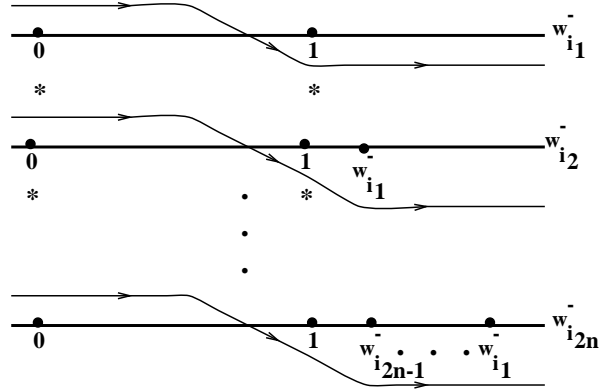,height=2in}}
\caption{Pictured are the positions of the
set of contours for {$w_i^-$} relative to the branch points
when the $w_i^+$ are between 0 and 1.}
\par
\end{figure}
The branch points marked by the *'s now appear below the contours as we
are performing the $w_{i_j}$ integrals in descending order of j.  The
placement of the $w^-_{i_j}$ branch points beyond 1 is justified by
the deformation of the contours that we now make as shown in Figure 5.
\begin{figure}
\par
\centerline{\psfig{figure=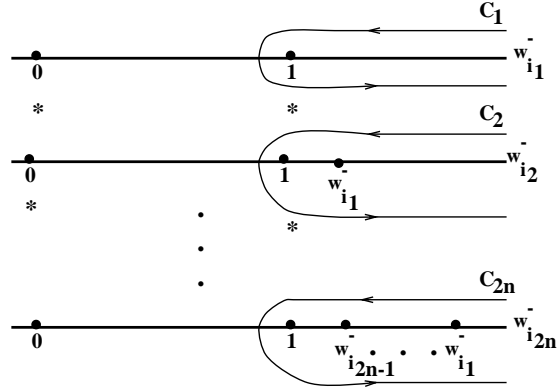,height=2in}}
\caption{Pictured are the deformations of the contours employed when
$w_1^+$ through $w_n^+$ are between 0 and 1.}
\par
\end{figure}
As we see, the contours are now such that the values of the
$w_i^-$'s are all greater than 1, thus justifying the placement of the
$w^-_i$ branch points beyond 1.  However, the given ordering of these
branch points should not be taken as fixed, but should be understood to
depend upon the specific values of the $w_i^-$'s.  For example, if we are
at a point on the contour $C_2$ such that $w^-_{i_2} > w^-_{i_1}$ then
for the contours $C_k$, $k>2$ the branch point for $w^-_{i_1}$ should
always fall to the right of that of $w^-_{i_2}$.  Thus, for example,
the ordering of
the branch points for the contour $C_{2n}$ as pictured in Figure 5 assumes
an integration region for the other 2n-1 contours where $w^-_{i_{2n-1}} <
\cdots < w^-_{i_1}$.

Now taking $\epsilon \rightarrow 0$, we thus are able to
express $I_{2n}$ as the following sum of integrals
\begin{equation}
I_{2n} = (-i)^{2n} |z|^{4n-2\alpha^2 - 2\hat\beta^2 n}
\sum_{\sigma \in S_{2n}} I \left( \sigma (1), \cdots ,\sigma
(2n) \right) ,
\end{equation}
\noindent where $S_{2n}$ is the permutation group of 2n objects and
\begin{equation}
I \left(i_1,\cdots ,i_{2n} \right) = J_1 \left( i_1, \cdots ,i_{2n}
\right) J_2 \left( i_1, \cdots ,i_{2n} \right),
\end{equation}
\noindent where $J_1$ and $J_2$ are given by
\vfill\eject

\begin{eqnarray} \label{j1}
J_1 \left( i_1, \cdots , i_{2n} \right) & = &
\int_0^1 \! dw_{i_1} w_{i_1}^{-\gamma_{i_1}} (1 - w_{i_1})^{\gamma_{i_1}}
\times \\
& & \nonumber \int_0^{w_{i_1}} \! dw_{i_2} w_{i_2}^{-\gamma_{i_2}}
(1 - w_{i_2})^{\gamma_{i_2}} (w_{i_1} - w_{i_2})^{\gamma_{i_1 i_2}}
\int_0^{w_{i_3}} \! dw_{i_3} \cdots \times \\
& & \nonumber \int_0^{w_{2n-1}} \! dw_{i_{2n}}
w_{i_{2n}}^{-\gamma_{i_{2n}}}
(1 - w_{i_{2n}})^{\gamma_{i_{2n}}}
(w_{i_1} - w_{i_{2n}})^{\gamma_{i_1i_{2n}}}
\cdots  \\
& & \nonumber \hspace{2.5in}
(w_{i_{2n-1}} - w_{i_{2n}})^{\gamma_{i_{2n-1} i_{2n}}}
\end{eqnarray}
\begin{eqnarray}
J_2 \left( i_1, \cdots , i_{2n} \right) & = &
\int_{C_1} \! dw_{i_1} w_{i_1}^{-\gamma_{i_1}} (1 - w_{i_1})^{\gamma_{i_1}}
\times \\
& & \nonumber \int_{C_2} \! dw_{i_2} w_{i_2}^{-\gamma_{i_2}}
(1 - w_{i_2})^{\gamma_{i_2}} (w_{i_1} - w_{i_2})^{\gamma_{i_1 i_2}}
\int_{C_3} \! dw_{i_3} \cdots \times \\
& & \nonumber \int_{C_{2n}} \! dw_{i_{2n}}
w_{i_{2n}}^{-\gamma_{i_{2n}}}
(1 - w_{i_{2n}})^{\gamma_{i_{2n}}}
(w_{i_1} - w_{i_{2n}})^{\gamma_{i_1 i_{2n}}} \cdots \times\\
& & \nonumber
\hspace{2.5in} (w_{i_{2n-1}} - w_{i_{2n}})^{\gamma_{i_{2n-1} i_{2n}}}
\end{eqnarray}
and we have defined the $\gamma$'s as follows
\begin{equation}
\hspace{-.45in}
\gamma_i = \left\{ \begin{array}{ll}
		{\alpha\hat\beta /2} & \mbox{$1 \leq i \leq n$} \\
		{-\alpha\hat\beta /2} & \mbox{$n < i \leq 2n$}
		\end{array}
	\right. ;
\end{equation}
\begin{equation} \hspace{1in}
\gamma_{ij} = \left\{ \begin{array}{ll}
	\hat\beta^2 & \mbox{$1 \leq i,j \leq n$ or $n < i,j \leq 2n$} \\
	-\hat\beta^2 & \mbox{{$1 \leq i \leq n \atop n < j \leq 2n$} or
			     {$1 \leq j \leq n \atop n < i \leq 2n$}}
			\end{array}
		\right. .
\end{equation}
In summing over the $J_1J_2$'s we rearrange the signs of the terms
involving powers of $w_i - w_j$.  But because we always do this in pairs,
one for $J_1$ and one for $J_2$, there are no phases between the terms in
the sum $\sum_{\sigma \in S_n}$.
We note that $I(i_1,\ldots ,i_{2n})$ is invariant under various
permutations of the $i_j$'s.
Any permutation not mixing values of the $i_j$'s between
1 through n with those n+1 through 2n leaves $I(i_1,\ldots ,i_{2n})$
unchanged.

Now $J_1$ is no more than a generalised Euler integral which we evaluate
in Appendix A.  The form is sufficiently complicated for the general case
that there is little point in writing it down here.  Indeed,
 \ref{j1}
offers a much
compact expression for $J_1$.  $J_2$, on the other hand, is as of yet
still a formal expression.  The contours $C_k$ enclose multiple branch
points and the corresponding cuts introduce phases.  These phases
depend upon which section of the integration region we are
in, as explained earlier.
These sections can be labelled by the specific order of the $w_i^-$'s
in the section.  As there are n! ways of ordering the $w_i^-$'s, $J_2$
is a sum of n! terms.
Now it is possible to develop a general notation for the assignment of
the phases in each of the n! cases.  However, this does not bring us any
farther in actually evaluating $I_{2n}$.  Thus it is at this point that
we specialize to specific values of n, in particular $n=1$.  This case,
together with the evaluation of $I_3$, brings out all the subtleties
inherit in the assignment of phases.

\begin{figure}
\par
\centerline{\psfig{figure=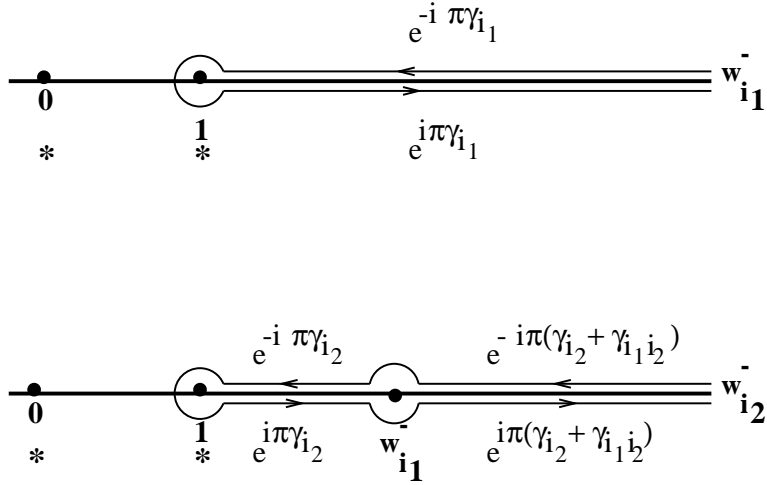,height=2.5in}}
\caption{Shown are the phase assignments for the integral $J_2(i_1,i_2)$.}
\par
\end{figure}

\subsection{Evaluation of $I_2$}
Assigning phases is particularly straightforward for the case of $I_2$.
 From Figure 5, n=1 yields phases as pictured in Figure 6.
$J_2 (i_1,i_2)$ then has the following form:
\nopagebreak
\begin{eqnarray}
J_2 (i_1,i_2) = & & 2 i s({\gamma_{i_1}}) \int^\infty_1 dw_{i_1}
(w_{i_1} - 1)^{\gamma_{i_1}} w_{i_1}^{-\gamma_{i_1}} \times \\
& & \nonumber \left[ 2 i s({\gamma_{i_2}}) \int^{w_{i_1}}_1 dw_{i_2}
(w_{i_2} - 1)^{\gamma_{i_2}} w_{i_2}^{-\gamma_{i_2}}
(w_{i_1} - w_{i_2})^{\gamma_{i_1 i_2}} + \right. \\
& & \nonumber \left. 2 i s({\gamma_{i_2} + \gamma_{i_1 i_2}})
\int_{w_{i_1}}^\infty dw_{i_2}
(w_{i_2} - 1)^{\gamma_{i_2}} w_{i_2}^{-\gamma_{i_2}}
(w_{i_1} - w_{i_2})^{\gamma_{i_1 i_2}} \right]
\end{eqnarray}
\noindent where we have introduced the notation $s(a) = \sin(\pi a)$.
In writing this expression for $J_2(i_1,i_2)$ down we have
assumed that the portions of the contours that circle
about the branch point contribute nothing.  This assumption can
always be made good by continuing the exponents of the terms in the
integrands to points where these portions do contribute nothing,
evaluating the integral, and then continuing back to the desired value of
the exponents in the final expression.  It is this continuation of the
exponents that makes the method akin to dimensional regularisation.  But
instead of continuing the dimension of space-time, we continue the
the parameter $\hat\beta$, i.e. the scaling dimension of the vertex
operators appearing in the action.

$J_2(i_1,i_2)$ is now in a form where it may be directly
evaluated.  Making the change of variables, $w_i \rightarrow 1 / w_i$,
the above may be written as
\begin{equation}
J_2(i_1,i_2) = -4 \left[s({\gamma_{i_1}})s({\gamma_{i_2}}) K(i_2,i_1)
+ s({\gamma_{i_1}})s({\gamma_{i_2} + \gamma_{i_1 i_2}}) K(i_1,i_2) \right] ,
\end{equation}
\noindent where $K(i,j)$ is defined to be
\begin{equation}
K(i,j) = \int^1_0 dw_i w_i^{-2 + \hat\beta^2}(1-w_i)^{\gamma_i}
	 \int^{w_i}_0 dw_j w_j^{-2 + \hat\beta^2}(1-w_j)^{\gamma_j}
	     (w_i - w_j)^{-\hat\beta^2},
\end{equation}
where we have used the fact that $\gamma_{12} = \hat\beta^2$ and that
$\gamma_1 = - \gamma_2$.  We evaluate $K(i,j)$ in Appendix B (even though
the K's are no more than Euler integrals, Appendix A is not sufficient to
evaluate the K's) with the result
\begin{eqnarray} \label{k12} \nn
K(i_1,i_2) & = & \B (1+\gamma_{i_1},\bh^2 -1)\Gamma(1-\bh^2)
\Gamma(\bh^2 - 1) (-\gamma_{i_2}) (\bh^2 -1) \\
& & \times \gh (1-\gamma_{i_2},\bh^2-1,\bhs,\bh^2+\gamma_{i_1},2,1) ,
\end{eqnarray}
\noindent where \gh ~is a standard hypergeometric function.

Using Appendix A to directly evaluate $J_1(i_1,i_2)$ we find
\begin{eqnarray} \label{j12}
\nn J_1 (i_1,i_2) & = &
\int^1_0 dw_{i_1} w_{i_1}^{-\gio} (1 - w_{i_1})^{\gio} \times \\
&& \hspace{.5in}
\int^{w_{i_1}}_0 dw_{i_2} w_{i_2}^{-\git}
(1 - w_{i_2})^{\git}(w_{i_1}-w_{i_2})^{\gamma_{12}} \\
\nn & = & B(1+\gio,2-\bhs)B(1-\bhs,1-\git) \times \\
& & \nonumber \hspace{.5in} \gh
(-\git,2-\bhs,1-\git,3-\bhs+\gio,2-\bhs-\git ,1).
\end{eqnarray}
\noindent So $I_2$ has the form

\begin{eqnarray} \label{two} \nonumber
I_2 \!\!\!& = & \!\!\! -4 |z|^{4-2\alpha^2 -2\hat\beta^2} \left[ J_1(1,2)
\left( s^2(\abot) K(2,1) + s(\abot) s(\bhs + \abot) K(1,2)
\right) \right. \\
& & \nonumber \hspace{1.15in} \left. + J_1(2,1)
\left( s^2(\abot) K(1,2) + s(\abot) s(\abot - \bhs) K(2,1)
\right) \right] . \\
& &
\end{eqnarray}

\subsection{Evaluation of $I_{2n+1}$}
We now move on the $I_{2n+1}$.  All of the same techniques that were used
to analyze $I_{2n}$ apply here.  We are thus able to immediately write
$I_{2n+1}$ as follows:
\begin{equation}
I_{2n+1} = (-i)^{2n+1} |z|^{2(2n+1)-\hat\beta^2(2n+{3\over 2})}
\sum_{\sigma \in S_{2n+1}} I \left( \sigma (1), \cdots ,\sigma
(2n+1) \right) ,
\end{equation}
\noindent where
$I = \left(i_1,\cdots ,i_{2n+1} \right) = J_1 \left( i_1, \cdots ,i_{2n+1}
\right) J_2 \left( i_1, \cdots ,i_{2n+1} \right),$
and $J_1$ and $J_2$ are given by
\begin{eqnarray}
J_1 \left( i_1, \cdots , i_{2n+1} \right) & = &
\int_0^1 \! dw_{i_1} w_{i_1}^{\de_{i_1}} (1 - w_{i_1})^{\de_{i_1}} \times
\\
& & \nonumber \int_0^{w_{i_1}} \! dw_{i_2} w_{i_2}^{\de_{i_2}}
(1 - w_{i_2})^{\de_{i_2}} (w_{i_1} - w_{i_2})^{\de_{i_1 i_2}}
\int_0^{w_{i_3}} \! dw_{i_3} \!\!\! \cdots \times \\
& & \nonumber \int_0^{w_{2n}} \!\!\! dw_{i_{2n+1}} \!\!\!
w_{i_{2n+1}}^{\de_{i_{2n+1}}}
(1 - w_{i_{2n+1}})^{\de_{i_{2n+1}}} (w_{i_1} \! -
\! w_{i_{2n+1}})^{\de_{i_1i_{2n+1}}} \\
& & \nonumber
\hspace{2.05in}
\cdots (w_{i_{2n}} - w_{i_{2n+1}})^{\de_{i_{2n} i_{2n+1}}}
\end{eqnarray}
\begin{eqnarray}
J_2 \left( i_1, \cdots , i_{2n} \right) & = &
\int_{C_1} \! dw_{i_1} w_{i_1}^{\de_{i_1}} (1 -
w_{i_1})^{\de_{i_1}} \times \\
& & \nonumber \int_{C_2} \! dw_{i_2} w_{i_2}^{\de_{i_2}}
(1 - w_{i_2})^{\de_{i_2}} (w_{i_1} - w_{i_2})^{\de_{i_1 i_2}}
\int_{C_3} \! dw_{i_3} \cdots \times \\
& & \nonumber \int_{C_{2n+1}} \! dw_{i_{2n+1}}
w_{i_{2n+1}}^{\de_{i_{2n+1}}}
(1 - w_{i_{2n}})^{\de_{i_{2n+1}}}
(w_{i_1} - w_{i_{2n+1}})^{\de_{i_1 i_{2n+1}}} \\
& & \nonumber \hspace{2.05in}
\cdots (w_{i_{2n}} - w_{i_{2n+1}})^{\de_{i_{2n} i_{2n+1}}}
\end{eqnarray}
and we have defined the $\de$'s as follows

\begin{equation}
\hspace{-.7in}
\de_i =  {\left\{ \begin{array}{ll}
                -\bhs /2 & \mbox{$1 \leq i \leq n+1$} \\
                \bhs & \mbox{$n+1 < i \leq 2n+1$}
                \end{array}
        \right.}
\end{equation}

\noindent and
\begin{equation}
 \hspace{.90in}
\de_{ij} = {\left\{ \begin{array}{ll}
        \hat\beta^2 & \mbox{$1 \leq i \neq j \leq n+1$ or
				$n+1 < i \neq j \leq 2n+1$} \\
        -\hat\beta^2 & \mbox{{$1 \leq i \leq n+1 \atop n+1 < j \leq 2n+1$}
						or
                             {$1 \leq j \leq n+1 \atop n+1 < i \leq 2n+1$}}
                        \end{array}
                \right.}
\end{equation}
\noindent Again $J_2$ is a formal object because as of yet phases have
not yet been assigned to the various portions of the contours $C_k$.  And
again there is no advantage of performing this assignment in general and
so we specialize to the cases of $I_1$ and $I_3$.

\subsection{Evaluation of $I_1$ and $I_3$}
Now $I_1$ is particularly simple.  The phase assignment is given in
Figure 7.

\begin{figure}
\par
\centerline{\psfig{figure=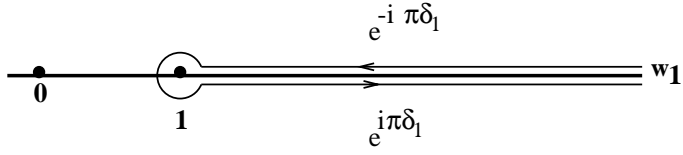,height=.75in}}
\caption{Shown are the phase assignments for the integral $J_2(1)$.}
\par
\end{figure}

\noindent Thus $J_2 (1)$ equals
\begin{eqnarray}
J_2 (1) & = & 2 i s(\delta_1) \int^\infty_1 dw_1 w_1^{\de_1}
(w_1 -1)^{\delta_1} \\
& = & \nonumber 2 i s(\delta_1) \int^1_0 dw_1 w_1^{-2-2\delta_1}
(1 - w_1)^{\delta_1}.
\end{eqnarray}

\noindent Now $J_1(1)$ and $J_2(1)$ are no more than the integral
representations of the beta-function.  Thus
\begin{eqnarray}
J_1(1) & = & \B (1+\delta_1,1+\delta_1) , \\
J_2(1) & = & 2i s(\delta_1) \B (-1-2\delta_1,1+\delta_1) ,
\end{eqnarray}
\noindent and so because $\delta_1 = -\bhs /2$
\begin{equation} \label{one}
I_1 = -2 |z|^{2-3\hat\beta^2/2} s(\bhst )
\B (1-{\bhs \over 2},1-{\bhs \over 2}) \B (\bhs - 1,1 - {\bhs \over 2}) .
\end{equation}
We mention that the integral $I_1$ is one that also
arises in calculating the 4-point
Virasoro-Sharpiro formula.
The methods used in this case for evaluating this
integral give the same answer as above.
\begin{figure}
\par
\centerline{\psfig{figure=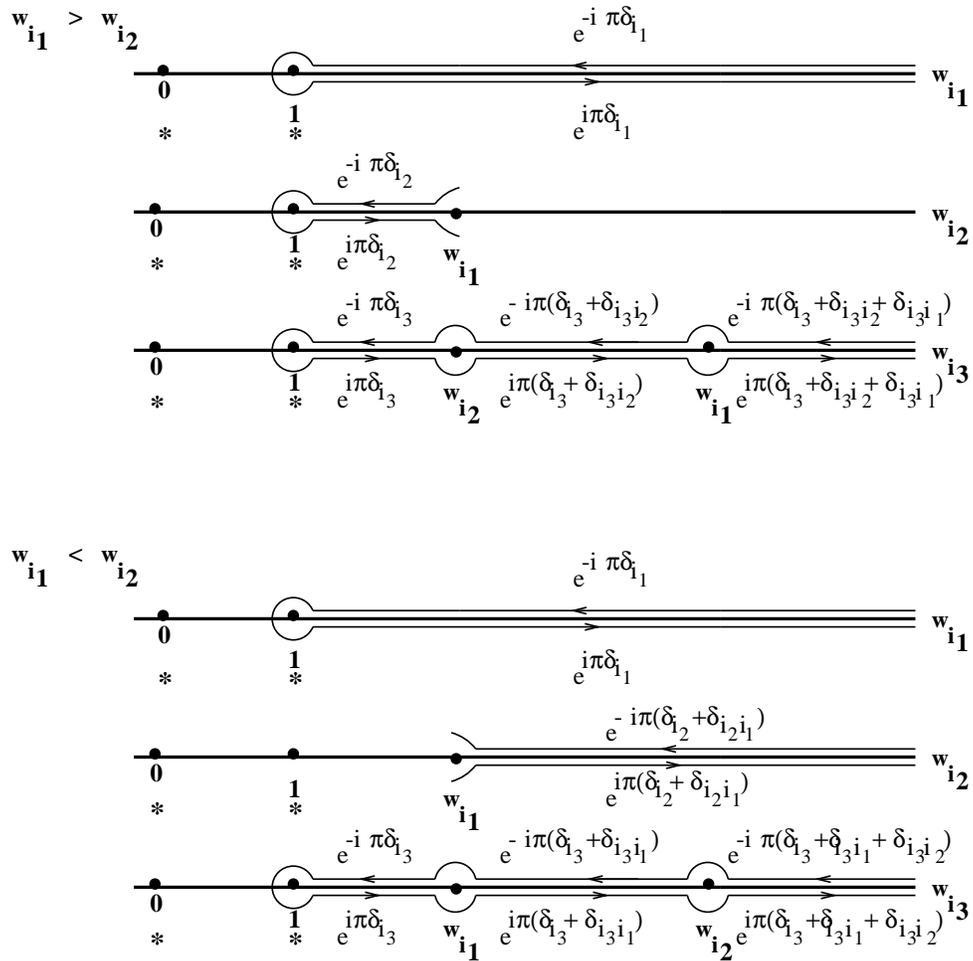,height=5.0in}}
\caption{Shown are the phase assignments for the integral
$J_2(i_1,i_2,i_3)$.}
\par
\end{figure}

The evaluation of $I_3$ is more complicated because the assignment of
phases is more complicated.  The assignment differs depending upon the
ordering of $w_{i_1}$ with $w_{i_2}$.  The phase assignment for the two
possible orderings is pictured in Figure 8.  Defining $K(i_1,i_2,i_3)$ by
\begin{eqnarray}
K(i_1,i_2,i_3) & = & \int^\infty_1 d\wi \wi^\di (\wi -1)^\di \times \\
& & \nonumber \int_1^\wi d\wii \wii^\dii (\wii - 1)^\dii (\wi -
\wii)^{\delta_{i_1}\delta_{i_2}} \times \\
& & \nonumber \int^\wii_1 d\wiii \wiii^\diii (\wiii -1)^\diii
(\wi - \wiii)^{\delta_{i_1i_3}} (\wii - \wiii)^{\delta_{i_2i_3}} ,
\end{eqnarray}
which after a change of variables $w_i \rightarrow 1/w_i$ can be written
as
\begin{eqnarray}
K(i_1,i_2,i_3) & = & \int^1_0 d\wiii
\wiii^{-2-2\diii -\delta_{i_3i_2} -\delta_{i_3i_1}} (1-\wiii)^\diii \\
& & \nonumber \int^\wiii_0 d\wii
\wii^{-2-2\dii -\delta_{i_2i_3} -\delta_{i_2i_1}} (1-\wii)^\dii
(\wiii - \wii)^{\delta_{i_3i_2}} \\
& & \nonumber \int^\wii_0 d\wi
\wi^{-2-2\di -\delta_{i_1i_3} -\delta_{i_1i_2}} (1-\wi)^\di
(\wiii - \wi)^{\delta_{i_3i_1}} \times \\
& & \nonumber \hspace{2.8in} (\wii - \wi)^{\delta_{i_1i_2}} ,
\end{eqnarray}
allows us to write $J_2(i_1,i_2,i_3)$ as
\vfill\eject
\begin{eqnarray}
J_2(i_1,i_2,i_3) = (2i)^3 \! \! \! & \! \! \! & \! \! \!
\left[s(\di )s(\dii ) s(\diii ) K(i_1,i_2,i_3) + \right. \\
& & \nonumber s(\di ) s(\dii ) s(\diii + \delta_{i_3i_2})
K(i_1,i_3,i_2) + \\
& & \nonumber s(\di ) s(\dii ) s(\diii +
\delta_{i_3i_2} + \delta_{i_3i_1})
K(i_3,i_1,i_2) + \\
& & \nn s(\di ) s(\dii +\delta_{i_1i_2}) s(\diii ) K(i_2,i_1,i_3) +\\
& & \nn s(\di ) s(\dii +\delta_{i_1i_2}) s(\diii + \delta_{i_3i_1})
K(i_2,i_3,i_1) + \\
& & \nn \left. s(\di ) s(\dii +\delta_{i_1i_2}) s(\diii + \delta_{i_3i_1}
+\delta_{i_3i_2})K(i_3,i_2,i_1) \right] .
\end{eqnarray}
Because $K_(i_1,i_2,i_3)$ is invariant under permutations of the i's
involving 1 and 2, it can easily be shown from the above expression that
\begin{eqnarray}
J_2(1,2,3) & = & 0; \\
J_2(1,3,2) & = & (2i)^3 s^2(\bhst ) (s(3\bhst ) + s(\bhst ))K(1,2,3); \\
\nn J_2(3,1,2) & = & (2i)^3 s(\bhst )s(3\bhst ) (s(3\bhst ) +s(\bhst
))K(1,2,3) + \\
& & s^2(\bhst ) (s(\bhst ) + s(3 \bhst ))K(1,3,2).
\end{eqnarray}
$J_2(1,2,3)$ vanishes indentically because of a cancellation of the phases.
Using the fact that both $J_1$ and $J_2$ possess the same symmetry as $K$,
$I_3$ then takes the form
\begin{eqnarray}
I_3 = 16 |z|^{6 - 7\hat\beta^2 /2} \! \! \! \! \!
& & \nn \left[ K(1,2,3) \left(
s^2(\bhst )(s(3\bhst ) + s(\bhst ))J_1(1,3,2) + \right. \right. \\
& & \nn \left. s(\bhst )s(3\bhst )
(s(3\bhst ) + s(\bhst ))J_1(3,1,2) \right) + \\
& & \nn \left.  K(1,3,2)
\left( s^2(\bhst ) (s(3\bhst ) + s(\bhst ))J_1(3,1,2)
\right) \right] . \\
\end{eqnarray}
Now all that remains is to evaluate the $J_1$'s and the $K$'s.

Using Appendix A we find
\begin{eqnarray}
\nn
K(1,2,3) \!\! & = & \!\! \B (1+\bhst ,-3+2\bhs ) \B (1-\bhs , -2+3\bhs )
\B (1+\bhs,\bhs - 1) \\
& & \nn \times \sum_{k_1k_2k_3}^\infty
{(\bhst )_{k_1} (\bhst )_{k_2} (\bhs )_{k_3} \over k_1! k_2! k_3!}
{(-3 +2\bhs )_{k_1+k_2} \over (-2+5\bhst )_{k_1+k_2}} \\
& & \label{k123}
\hspace{1.0in} \times {(-2 + 3\bhs )_{k_1+k_2+k_3} \over (-1 +
2\bhs)_{k_1+k_2+k_3}}
{(-1 + \bhs )_{k_2+k_3} \over (2\bhs )_{k_2+k_3}} ;\\
\nn K(1,3,2) \!\! & = & \!\! \B (1-\bhst ,-3+2\bhs ) \Gamma (1-\bhs)
\Gamma (-2+\bhs ) \Gamma (1-\bhs) \Gamma (\bhs - 1) \\
& & \nn \times \sum_{k_1k_2k_3}^\infty
{(-\bhst )_{k_1} (\bhst )_{k_2} (-\bhs )_{k_3} \over k_1! k_2! k_3!}
{(-3 +2\bhs )_{k_1+k_2} \over (-2+3\bhst )_{k_1+k_2}} \\
& & \hspace{1.0in}
\times {(-2 + \bhs )_{k_1+k_2+k_3} \over \Gamma (-1+k_1+k_2+k_3)}
{(-1 + \bhs )_{k_2+k_3} \over \Gamma (k_2+k_3)} ;\\
\nn J_1(1,3,2) \!\! & = & \!\! \B (1-\bhst ,3-3\bhst ) \B (1-\bhs ,2-\bhs )
\B (1-\bhs,1-\bhst ) \\
& & \nn \times \sum_{k_1k_2k_3}^\infty
{(-\bhst )_{k_1} (\bhst )_{k_2} (-\bhs )_{k_3} \over k_1! k_2! k_3!}
{(3 -3\bhst )_{k_1+k_2} \over (4-2\bhs )_{k_1+k_2}} \\
& & \hspace{1.0in} \times {(2 - \bhs )_{k_1+k_2+k_3} \over (3 -
2\bhs)_{k_1+k_2+k_3}}
{(1 - \bhs/2 )_{k_2+k_3} \over (2-3\bhst )_{k_2+k_3}} ;\\
\nn J_1(3,1,2) \!\! & = & \!\! \B (1+\bhst ,3-3\bhst ) \B (1-\bhs ,2)
\B (1+\bhs,1-\bhst ) \\
& & \nn \times \sum_{k_1k_2k_3}^\infty
{(\bhst )_{k_1} (\bhst )_{k_2} (\bhs )_{k_3} \over k_1! k_2! k_3!}
{(3 -3\bhst )_{k_1+k_2} \over (4-\bhs )_{k_1+k_2}} \\
& & \label{j312} \hspace{1.0in}
\times {(2)_{k_1+k_2+k_3} \over (3-\bhs)_{k_1+k_2+k_3}}
{(1-\bhs/2 )_{k_2+k_3} \over (2+\bhst )_{k_2+k_3}} .
\end{eqnarray}

\noindent In evaluating
these integrals the same sort of techniques that went into
evaluating $K(i_1,i_2)$ (i.e. handling the appearance of a pole
multiplying a zero - see Appendix B) were employed.  This completes the
evaluation of $I_3$.

\subsection{Evaluation of the Bubbles}

We now examine the contribution of the bubble diagrams (i.e. the
integrals in the denominator of the perturbative expansion).  We will show
that with our techniques for evaluating the integrals, the bubbles vanish
identically.  The integrals that we must evaluate are
\begin{equation}
I^B_{2n}  = \int d^2w_1 \cdots d^2w_{2n}
\prod_{1 \leq i \not= j \leq n \atop 2n \geq i \ne j > n}
(|w_i - w_j|^2)^{\hat\beta^2}
\prod_{1 \leq i \leq n < j \leq 2n} (|w_i - w_j|^2)^{-\hat \beta^2}.
\end{equation}
Performing the change of variables $x_i \rightarrow -ie^{-i2\epsilon}x_i$,
$w_i^\pm = t_i \pm x_i$ as before, $I^B_{2n}$ becomes
\begin{eqnarray}
\nn I_{2n}^B & = &
(-i)^{2n} \int \! dw^+_i dw^-_i
\eta (w^+_1,\cdots,w^+_{2n},\epsilon)
\eta (w^-_1,\cdots,w^-_{2n},-\epsilon),
\end{eqnarray}
where $\eta$ takes the from
\begin{equation}
\eta (w_1, \cdots ,w_{2n}, \epsilon) = \prod^{2n}_{i=1} \prod^n_{j=i+1}
(w_i - w_j - i\epsilon (\Delta_i - \Delta_j) )^{\pm \hat\beta^2} ,
\end{equation}
where the sign of the exponents deponds upon the particular values of $i$
and $j$.  Again taking the ordering of the $w^+_i$'s to be
$ w^+_{i_1} > \cdots w^+_{i_{2n}}$, the deformation of the $w_i^-$
contours about the branch points the appear as in Figure 9.

\begin{figure}
\par
\centerline{\psfig{figure=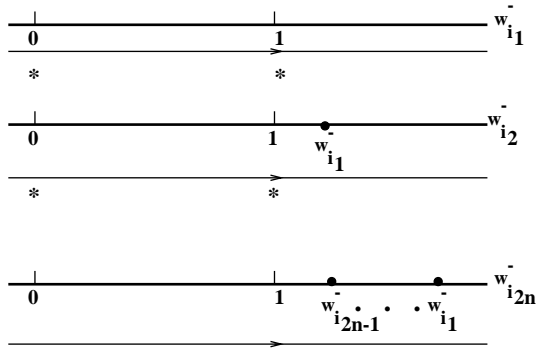,height=1.75in}}
\caption{Pictured are the set of contours for {$w^-_{i_1}$} arising from
the integration of the integral $I^B_{2n}$.  In contrast to the situation
for $I_{2n}$ and $I_{2n+1}$, there are no branch points at 0 and 1.}
\par
\end{figure}

Because there are no branch points at 0 and 1, the $w^-_{i_1}$-contour
(or, indeed, the $w^-_{i_{2n}}$-contour) can always
be closed at $\infty$.  Thus $I_{2n}^B$ is identically 0.
In Figure 9 a specific ordering of the $w^-_i$ branch points has been
assumed.  However, changing this
ordering does not change the argument just made.
The vanishing of the bubbles is
supported by our analysis of the Ising spin correlators.
Because our results match previous calculations, it is likely that
we have accurately taken into account the perturbative structure of the
vacuum.

\vfill\eject
\subsection{Summary of Results}
At this point we gather previous results to provide expressions for
$G (\alpha ,-\alpha )$ and $G (\hat\beta /2,\hat\beta /2)$:
\begin{eqnarray}
G (\alpha ,-\alpha ) & = & |z|^{-2\alpha^2} +
\left({\lambda \over 2\pi} \right)^2 I_2 \\
& = & \nn |z|^{-2\alpha^2} -
\left({\lambda \over \pi} \right)^2
|z|^{4-2\alpha^2-2\hat\beta^2} \times \\
& & \nn \hspace{.50in} \left[ J_1(1,2)
\left( s^2(\abot) K(2,1) + s(\abot) s(\bhs + \abot) K(1,2)
\right) \right. + \\
& & \nn \hspace{.50in} J_1(2,1)
\left. \left( s^2(\abot) K(1,2) + s(\abot) s(\abot - \bhs) K(2,1)
\right) \right] ;
\end{eqnarray}
\vspace{-.1in}
\begin{eqnarray}
G ({\hat\beta \over 2},{\hat\beta \over 2}) & = &
\left({\lambda \over 2\pi} \right) I_1 +
{1 \over 2} \left({\lambda \over 2\pi} \right)^3 I_3 \\
& = & \nn -{\lambda \over \pi}
|z|^{2-3\hat\beta^2/2} s(\bhst )
\B (1-\bhst ,1-\bhst) \B (\bhs -1, 1-\bhst) \\
& & \nn + \left({\lambda \over \pi} \right)^3 |z|^{6-7\hat\beta^2/2}
\times \\
& & \nn \hspace{.7in} \left[ K(1,2,3) \left(
s^2(\bhst )(s(3\bhst ) + s(\bhst ))J_1(1,3,2) + \right. \right. \\
& & \nn \hspace{.7in} \left. s(\bhst )s(3\bhst )
(s(3\bhst ) + s(\bhst ))J_1(3,1,2) \right) + \\
& & \nn \hspace{.7in} \left.
K(1,3,2) \left( s^2(\bhst ) (s(3\bhst ) + s(\bhst ))J_1(3,1,2)
\right) \right].
\end{eqnarray}
The expressions for $J(i,j)$ and $K(i,j)$ are given in \ref{k12} and
\ref{j12} respectively.
The expressions for $J(i,j,k)$ and $K(i,j,k)$ are given in
\ref{k123} through \ref{j312}.

\setcounter{equation}{0}

\section{Sine-Gordon as Ising and SU(2) Gross-Neveu}
In this section we examine our expressions in the limits in which the
sine-Gordon theory maps onto a doubled Ising model ($\hat\beta = 1$) and
the Gross-Neveu SU(2) model ($\hat\beta = \sqrt{2}$).  We first examine the
doubled Ising model.

\subsection{Doubled Ising Model}
As demonstrated in section 3, as \bh ~ approaches 1, the perturbative
expansion develops IR singularities.  In our expressions, these
singularities manifest themselves as poles in gamma functions.  These
poles are indicative of the logs that appear in the Ising
spin-correlators in the scaling limit (see, for example, \cite{wu}).
Specifically, a n-th order pole term indicates a n-th order log.  The logs
are obtained by expanding the powers of $|z|$ multiplying the pole terms.
For example, writing $\bhs = 1 + \epsilon$ with $\epsilon$ small,
a typical situation will see the following expansion:
\begin{equation}
{|z|^{1-\hat\beta^2} \over \bhs - 1} = {1 \over \epsilon} - \log|z| + \cdots
{}~.
\end{equation}
After this expansion is made, the divergent pieces (i.e. the $1/\epsilon$
term) are subtracted away.  This can be understood in analogy with a
cut-off method of regularisation.  Throwing away the divergent terms
amounts to no more than a specific choice of the cutoff.
This procedure differs from that used by
Dotsenko \cite{dot} where the powers of $|z|$ are not expanded and
logs are substituted directly for the $1 / \epsilon $ terms.  This procedure
is equivalent to ours in situations involving first order logs.
However it fails to produce the
correct coefficients of higher order logs that arise from overlapping
divergences.

To test this regularisation
technique, we examine a correlator simpler than the
Ising correlator, the free-fermion correlator.
As such we will need to review some aspects of the connection between the
free-fermion and the sine-Gordon theory at $\bh = 1$.  The free fermion
action in Euclidean space-time reads
\begin{equation}
S = {1 \over 4\pi} \int dx dt \left[ \bar\psi_- \partial_z \bar\psi_+ +
\psi_-\partial_{\bar z} \psi_+ + im (\psi_-\bar\psi_+
- \bar\psi_-\psi_+)\right] ,
\end{equation}
where we have introduced the Dirac spinors $\Psi_{\pm} =
\left( {\bar\psi_{\pm} \atop \psi_{\pm}} \right)$ which satisfy
$\Psi_{+} = \Psi_-^\dagger$ and employed
an appropriate choice
of gamma matrices
\footnote{~$\gamma^0 = \left({0 \atop i} {-i \atop 0} \right)$,
$\gamma^1 = \left({0 \atop -i} {-i \atop 0} \right).$}.
The equations of motion for these fields are
\begin{eqnarray}
\nn \partial_z\bar\psi_{\pm} & = & i m \psi_{\pm} , \\
& & \\
\nn \partial_{\bar z}\psi_{\pm} & = & -i m \bar\psi_{\pm}.
\end{eqnarray}

The correlators we will examine are
\begin{equation}
\langle\bar\psi_-(z,\bar z) \psi_+(0) \rangle ~~ \rm{and} ~~
\langle\psi_-(z,\bar z) \psi_+(0) \rangle .
\end{equation}
Knowing the first of these correlators, the equations of motion allow us
to fix the second.  With this in mind, the on-shell mode expansions are:
\begin{eqnarray}
\nn \psi_{\pm} & = & \pm \sqrt{m} \int^{\infty}_{-\infty}
{d\theta \over 2\pi i}
e^{\theta /2} \left( c^{\mp}(\theta)e^{\mp i p(\theta)\cdot x} -
d^{\pm}(\theta) e^{\pm i p(\theta)\cdot x} \right) \\
& & \\
\nn \bar\psi_{\pm} & = & -i \sqrt{m} \int^{\infty}_{-\infty} {d\theta \over
2\pi i}
e^{-\theta /2} \left( c^{\mp}(\theta)e^{\mp i p(\theta)\cdot x} +
d^{\pm}(\theta) e^{\pm i p(\theta)\cdot x} \right).
\end{eqnarray}
In writing these expansions we have introduce the rapidity variable
$\theta$ by which on-shell Euclidean momentum-energy is parameterized,
\begin{equation}
\nn E = im \cosh (\theta) ~~ \rm{and} ~~ p = m \sinh (\theta),
\end{equation}
so that $-E^2 - p^2 = m^2$ and $ip(\theta)\cdot x = m(ze^\theta + \bar
z e^{-\theta} )$.  The creation (+) and destruction
(-) operators obey the following commutation relations
\begin{equation}
\{ d^-(\theta), d^+(\theta ') \} = \{ c^-(\theta), c^+(\theta ') \} =
4 \pi^2 \delta(\theta - \theta ').
\end{equation}
Using these mode expansions together with the equations of motion,
it is then straightforward to calculate the
two correlators
\begin{eqnarray}
\langle \bar\psi_- (z,\bar z) \psi_+ (0) \rangle & = & -2 i m \rm{K}_0
(mr), \\
\label{cor}
\langle \psi_- (z,\bar z) \psi_+ (0) \rangle & = & 2 m
\sqrt{\bar z \over z} \rm{K}_1 (mr).
\end{eqnarray}
where the $\rm{K}$'s are the standard modified Bessel functions.

The relavence of these correlators is found in the demonstration of
Mandlestam \cite{mand} that the spinor
components can be expressed in terms of the quasi-chiral components of
the sine-Gordon field, $\phi^L$ and $\phi^R$:
\begin{eqnarray}
\nn \psi_{\pm} & = & \exp(\pm i \phi^L), \\
& & \\
\nn \bar\psi_{\pm} & = &  \exp(\mp i \phi^R),
\end{eqnarray}
The fields $\phi^L$ and $\phi^R$ are termed quasi-chiral because in the
massless limit,
\begin{equation}
\partial_{\bar z} \phi^L = \partial_{z} \phi^R = 0 .
\end{equation}
In this limit $\phi^L$ is the left mover and $\phi^R$ the right mover.

With these conventions we can make the following identification
\begin{equation}
\label{fer}
\langle e^{i\phi^R( \bar z )} e^{i\phi^L (0)} \rangle = -2 i m \rm{K}_0(mr) .
\end{equation}
At first order the l.h.s. of this equation reduces to
\begin{eqnarray} \label{first}
\nn F( \hat\beta = 1) & \equiv & {\lambda \over 2 \pi} \int d^2w
\langle e^{i\phi^R(\bar z)} e^{i\phi^L(0)}
e^{-i\Phi(w,\bar w)}\rangle_{CFT} \\
& = & -i{\lambda \over 2\pi} \int
d^2w {1 \over (\bar w - \bar z)} {1 \over w}.
\end{eqnarray}
The $i$ arises from taking apart $e^{i\Phi}$, i.e.
$e^{\Phi (w,\bar w )} = -i e^{-i\phi^L (w)}e^{-i\phi^R (\bar w )}$.

This integral has a logarithmic divergence.  To regulate it we continue
the scaling dimensions of the operators (via continuing $\hat\beta$)
as follows:
\begin{eqnarray}
\nn F(\hat\beta) & = & {\lambda \over 2 \pi} \int d^2w \langle
e^{i\hat\beta\phi^R(\bar z)}
e^{i\hat\beta\phi^L(0)}
e^{-i\hat\beta\Phi(w,\bar w)} \rangle \\
& = & -i {\lambda \over 2\pi } \int
d^2w {1 \over (\bar w - \bar z)^{\hat\beta^2}}
{1 \over w^{\hat\beta^2}}.
\end{eqnarray}
Notice that the U(1) charge of the vertex operators still sums to zero.
Writing this integral in terms of Euclidean x-t coordinates, we easily
evaluate it to be
\begin{equation}
F(\hat\beta) = {2 i \lambda \over \pi} |z|^{2 - 2\hat\beta^2}
{s(2 - 2\bhs ) \over 2 - 2\bhs} \B (2\bhs - 2, 2 - \bhs) .
\end{equation}

We now intend to take $\hat\beta^2$ to $1$ via $\hat\beta^2 = 1+\epsilon$.
$\lambda$ then has scaling dimensions of $1 - \epsilon$.  So we may
express $\lambda$ as $\lambda = -m\mu^{-\epsilon}$ where $\mu$ is a mass
scale fixed to be
\begin{equation}
\mu^{-\epsilon} = (1-\epsilon\gamma + O(\epsilon^2))m^{-\epsilon}
\end{equation}
by Zamolodchikov's $\lambda - m$ relationship (see \ref{lamm}).  Taking
$\epsilon \rightarrow 0$ in $F(\hat\beta )$ then gives us
\begin{eqnarray}
F(1+\ep ) & = & -{i \over \epsilon} m \mu^{\epsilon}
|z\mu|^{-2\epsilon} \\
& = & - i m\mu^{\ep}\left( {1 \over \epsilon} - 2\log (m|z|) - 2\gamma
\right)
\end{eqnarray}
As indicated before we drop the $1/\epsilon$ term.
The above expression then reduces to
\begin{equation}
F(1) = 2 i m \left( \log ({mr \over 2}) + \gamma \right)
\end{equation}
as $|z| = r/2$.  In taking the limit $\epsilon \rightarrow 0$ we have not
expanded out the $\mu^{\ep}$ term so as to preserve the scaling
dimension of $F(\hat\beta )$.
Expanding out $\rm{K}_0(m r)$ to first order in $m$ shows
the two sides of \ref{fer} to be in agreement.  We thus have gained
some confidence
that our methods of evaluating the perturbative terms are correct.

To illustrate the connection between our method of regularisation and a
cut-off method of regularisation, considering regulating $F(1)$ with a
cut-off $|w| <R$:
\begin{equation}
F(1) = -i{\lambda \over 2\pi} \int^R d^2w {1 \over (\bar w - \bar z)}
{1 \over w}.
\end{equation}
Doing the integral we find
\begin{equation}
F(1) = 2 i \lambda \log ({R \over |z|}) + const.
\end{equation}
Now $R$ can be written as $R = abm^{-1}$ where a and b are dimensionless
constants and $a$ is to be taken to $\infty$ (as the cutoff is removed).
So as $\lambda = -m$
\begin{equation}
F(1) = 2 i m \log (a^{-1}b^{-1}m|z|) + const.
\end{equation}
To obtain a match between $F(1)$ and the known form of $\langle \bar
\psi_-(z, \bar z) \psi_+ (0) \rangle$ $\log (a)$ must be thrown away and
a specific choice for $b$ must be made.  The throwing away of $\log (a)$
corresponds to our discarding $1/\ep$.  However in our method,
to our advantage, there
is no need to fix the arbitrary constant, $b$.

The fermion correlators also provide an opportunity to test whether we
are accurately taking into account the vacuum structure of the theory.
The correlator $\langle\psi_- (z,\bar z) \psi_+ (0) \rangle$ has its first
non-trivial contribution at second order in $\lambda$, the first order at
which the bubbles contribute.  Bosonizing this corrrelator and expanding
to second order we obtain
\begin{eqnarray} \label{tcor}
2m\sqrt{\bar z \over z} \rm{K_1}(mr) & = &
\langle \psi_- (z,\bar z) \psi_+ (0) \rangle
\\
\nn & = & \langle e^{-i\phi^L (z, \bar z)} e^{i\phi^L (0)} \rangle \\
\nn & = & {1 \over z} + {1 \over 4\pi^2} K (1) + O(\lambda^4) ,
\end{eqnarray}
where
\begin{equation}
\label{kdiv}
K(1) = {\lambda^2 \over z} \int d^2w_1 d^2w_2 w_1 w_2^{-1}
(w_1 - z)^{-1} (w_2 - z) |w_1 - w_2|^{-2} .
\end{equation}
We note that the zeroth order term of this correlator agrees with
expansion of $\rm{K}_1$.

The integral $K$ in \ref{kdiv} is divergent.  (In fact it is quadratically
divergent.  If the bubbles were explicitly included we would only be
facing a logarithmic divergence.  That our evaluation of $K$ still leads
only to log's indicates we have correctly handled the bubbles.)
Regulating $K(1)$ as before by continuing $\hat\beta$ we obtain
\begin{equation}
K(\hat\beta) = {\lambda^2 |z|^{4-2\hat\beta^2} \over z^{\hat\beta^2}}
\int d^2w_1 d^2w_2 w_1^{\hat\beta^2} w_2^{-\hat\beta^2}
(w_1 - 1)^{-\hat\beta^2}
(w_2 - 1)^{\hat\beta^2} |w_1 - w_2|^{-2\hat\beta^2} .
\end{equation}
Using the techniques described in section 4, $K(\hat\beta )$ is easily
evaluated to be
\begin{eqnarray}
K(\bh) & = & -4\lambda^2 |z|^{4-2\hat\beta^2} z^{-\hat\beta^2} \times \\
\nn & & \hspace{.20in} \left[
K_1(\bh ) \left( s^2(\bhs )K_2(\bh ) + \left(s(\bhs )s(2\bhs ) + s^2(\bhs
) \right) K_2(-\bh ) \right) \right]
\end{eqnarray}
where $K_1(\bh )$ and $K_2(\pm \bh )$ equal
\begin{eqnarray}
K_1(\bh ) & = & \int^1_0 dw_1 \int^{w_1}_0 dw_2 (w_1 - w_2)^{-\hat\beta^2}
= {\rm{B} (1-\bhs , 1) \over 2 - \bhs} \\
\nn K_2(\pm \bh ) & = & \int^1_0 dw_1 w_1^{\hat\beta^2 -2}
(1-w_1)^{\mp\hat\beta^2} \int^{w_1}_0 dw_2 w_2^{\hat\beta^2-2}
(1-w_2)^{\pm\hat\beta^2}(w_1 - w_2)^{-\hat\beta^2} \\
\nn & = & \bhs (1-\bhs ) \Gamma(\bhs - 1, 1-\bhs ) \rm{B} (\bhs
-1,1\mp\bhs ) \\
&& \nn \hspace{1.2in} \gh (1\mp\bhs , \bhs , \bhs -1 , 2 ,\bhs \mp \bhs , 1)
\\
\nn & = & \bhs (1-\bhs) \Gamma^2 (\bhs - 1) \Gamma(1-\bhs )
\Gamma(2-\bhs ) \\
&& \hspace{1.2in} \gh (1\pm\bhs , \bhs - 1,2-\bhs ,2,1,1),
\end{eqnarray}
where in the last line we have used the analytic continuation formula in
appendix C.
Now writing $\bhs = 1 + \ep$ and taking $\ep \rightarrow 0$ we find
\begin{eqnarray}
K_1(\bh ) & = & -{1 \over \ep} \left( 1 + \ep + O(\ep^2) \right) \\
K_2(\bh ) & = & {1 \over \ep^2} \left( \ep + O(\ep^2) \right) \\
K_2(-\bh ) & = & {1 \over \ep^2} \left( 1 + \ep + O(\ep^2) \right).
\end{eqnarray}
\noindent Some of the
details of this calculation may be found in appendix D.
Combining
these results we find $K(1+\ep )$ to be
\begin{eqnarray}
\nn K(1+\ep ) & = & -4\pi^2 |mz|^2 |\mu z|^{2-2\hat\beta^2} z^{-\hat\beta^2}
{1\over\ep} (1+\ep ) \\
& = & 4\pi^2 |mz|^2 z^{-\hat\beta^2} \left( -{1\over\ep} + 2\left(
\log ({mr \over 2}) + \gamma - {1\over 2} \right) \right)
\end{eqnarray}
\noindent where we have used
$\mu^{-\ep} = \left( 1-\ep\gamma + O(\ep^2)\right) m^{-\ep}$
and $|z| = r/2$. We
again preserve the correct scaling dimensions of $K(\bh )$ by not
expanding $z^{-\hat\beta^2}$.  Instead we expand only
$|z|^{4-2\hat\beta^2}$, the z-dependence scaled out from the integral.
It is this piece, coming from the integral, that we expect to generate the
log's.  Dropping the infinite piece in $K(1+\ep )$,
we find for $\langle \psi_-(z, \bar z) \psi_+ (0) \rangle$
\begin{eqnarray}
\nonumber \langle \psi_-(z, \bar z) \psi_+ (0) \rangle & = & {1 \over z} +
{2 \over z} |mz|^2 \left( \log ({mr \over 2}) + \gamma - {1 \over 2}
\right) + O(m^4) \\
\nn & = & 2m \sqrt{\bar z \over z} \left( {1\over mr} + {mr \over 2} \left(
\log ({mr \over 2}) + \gamma - {1\over 2} \right) \right) + O(m^4). \\
\end{eqnarray}
\noindent Expanding out $K_1(mr)$,
we find the two sides of \ref{tcor} agree.  Thus
the bubble diagrams, even though not explicitly included, are taken into
account.

Having demonstrated that our regularisation techniques work on fermion
correlators, we now apply the same methods on the
Ising spin correlator with some
confidence.  We recall the spin correlator is given by
\begin{equation}
\langle \sigma(z,\bar z) \sigma (0) \rangle^2 = {1\over 2}
\left( G({1\over 2},-{1\over 2}) - G({1\over 2},{1 \over 2}) \right) .
\end{equation}
 From Wu et al. \cite{wu}, the expansion in the scaling limit of the spin
correlator is
\begin{equation} \label{scale}
\langle \sigma(z,\bar z) \sigma (0) \rangle^2 =
{1 \over R^{1/2}} \left( 1 + t\Omega + t^2({\Omega^2 \over 4} + {1 \over
8}) + {t^3 \over 8} \Omega + \cdots \right),
\end{equation}
where $\Omega = \log ({t \over 8}) + \gamma$, $R \propto r$, and t, the
scaling variable, is proportional to $mr$.

We can use the zeroth order terms on both sides of \ref{scale} to fix $R$
in terms of $r$.  We find
\begin{equation}
R = 2 r
\end{equation}
At the next order, \ref{scale} reduces to
\begin{equation}
{t \Omega \over R^{1/2}} = {1 \over 4\pi} L(1),
\end{equation}
were $L(1)$ equals
\begin{eqnarray}
\nn L(1) & \equiv &
-\lambda \int d^2w \langle e^{i\Phi (z)/2} e^{i\Phi (0)} e^{-i\Phi (w)}
\rangle \\
& = & -\lambda \int d^2w |z|^{1/2} |w-z|^{-1} |w|^{-1}.
\end{eqnarray}
As with the first order term of the fermion correlator $\langle \bar
\psi_-(z, \bar z) \psi_+(0) \rangle$, this integral has a logarithmic
divergence.  Regulating $L$ as before by taking
$L(1) \rightarrow L(\bh )$, we find
\begin{equation}
L(\bh ) = -\lambda |z|^{\eb /2} |z|^{2-2\eb} \int d^2w |w-1|^{-\eb}
|w|^{-\eb}.
\end{equation}
This integral has been done previously (see \ref{one}) with the result
\begin{equation}
L(\bh ) = 2 \lambda |z|^{\eb /2} |z|^{2-2\eb} s(\bhs /2) \B (1-\bhs /2 ,
1-\bhs /2 ) \B (1-\bhs /2,\bhs - 1).
\end{equation}
Taking $\bhs$ to 1 via $\bhs = 1 + \ep$ we find
\begin{equation}
L(1+\ep ) = -2\pi m |z|^{\eb /2} \mu^{\eb - 1} |\mu z|^{2 - 2\eb}
\left( {1 \over \ep} + 4 \log (2) \right).
\end{equation}
As before we do not expand $|z|^{\eb /2}\mu^{\eb - 1}$ so as to preserve
the correct scaling dimension.  We then have
\begin{equation}
L(1) = 4\pi m |z|^{1/2} (\log ({mr \over 8}) + \gamma ).
\end{equation}
Taking the scaling variable, $t$, to equal $mr$ we see $L(1) = {4\pi t
\over R^{1/2}} \Omega$.  Comparing this with \ref{scale} we see we have
agreement.

At second order, the perturbative expansion of the spin correlator gives
\begin{equation} \label{above}
{1 \over R^{1/2}} t^2 \left( {\Omega^2 \over 4} + {1 \over 8} \right) =
{1 \over 8\pi^2} M(1)
\end{equation}
where $M(1)$ equals
\begin{eqnarray}
\nn M(1) & = & \lambda^2 \int d^2w d^2y
\langle e^{i\Phi (z)/2} e^{-i\Phi (0)/2}
e^{i\Phi (w)}e^{i\Phi (y)} \rangle \\
& = & \lambda^2 |z|^{3/2} \int d^2w d^2y |w-y|^{-2}
|w-1| |w|^{-1} |y-1|^{-1} |y|.
\end{eqnarray}
We will show \ref{above} is valid to the leading log term.

$M(1)$ is (quadratically) divergent.  Again, as with the second order
term of the fermion correlator, explicit inclusion of the bubble
contribution leaves $M(1)$ logarithmically divergent.  But, as before,
explicit inclusion of the bubbles is unnecessary.  To regulate, we
continue $\bh$.  $M(\bh )$ is then
\begin{equation}
M(\bh ) =  \lambda^2 |z|^{4- 5\eb /2} \int d^2w d^2y |w-y|^{-2\eb}
|w-1|^{\eb} |w|^{-\eb} |y-1|^{-\eb} |y|^{\eb}.
\end{equation}
We have already evaluated this integral.  Copying the result in \ref{two}
gives us
\begin{eqnarray} \nn
M(\bh ) & = &  -4 \lambda^2 |z|^{4-5\eb /2} \left[ J_+
\left( s^2(\bhs /2) K_- + s(\bhs /2) s(3\bhs /2) K_+
\right) \right. \\
& & \hspace{.9in} \left. + J_-
\left( s^2(\bhs /2) K_+ - s^2(\bhs /2) K_-
\right) \right] .
\end{eqnarray}
where
\begin{eqnarray}
K_{\pm} (\bh ) & = &\pm \B ( 1\pm \bhs/2 , \bhs -1) \Gamma(1-\bhs )
\Gamma(\bhs - 1)
({\bhs \over 2}) (\bhs - 1) \times \\
\nn & & \hspace{.15in} \gh (1\pm\bhs /2,\bhs ,\bhs -1,\bhs\pm\bhs
/2,2,1) \\
J_{\pm} (\bh) & = & \B (1\pm \bhs /2,2-\bhs ) \B (1-\bhs , 1\pm\bhs /2)
\times \\
\nn & & \hspace{.15in} \gh (\pm
\bhs /2,2-\bhs ,1\pm\bhs /2,3-\bhs \pm\bhs /2,2-\bhs\pm\bhs /2,1).
\end{eqnarray}
Setting $\bhs = 1+\ep$ we find
\begin{eqnarray}
K_{\pm}(\ep) & = & \mp {1 \over 2\ep^2}
\left[ 1 + \ep (2 - \gamma - \psi(1\pm 1/2) ) + O(\ep^2) \right] \\
J_{\pm}(\ep) & = & -{1\over\ep} \left[1 + \ep(\psi(1\pm 1/2) + \gamma \mp
1 \pm \pi^2/4 ) + O(\ep^2) \right]
\end{eqnarray}
where $\psi$ is the logarithmic derivative of the gamma-function, i.e.
$\psi(x) = \Gamma '(x)/\Gamma (x)$.  The details of this calculation may
be found in appendix E.

Putting everything together we find for $M(1+\ep )$
\begin{eqnarray}
M(1+\ep ) & = & -4 |z|^{4-5\eb /2} \lambda^2 {1 \over \ep^2} \left(
-{\pi^2 \over 2} + O(\ep^2 ) \right) \\
& = & \nn {2\pi^2 \over \ep^2} |z|^{-\eb /2} |zm|^2
\left( 1 - 2\ep\log (z\mu) +2\ep^2\log^2(z\mu) \right)
\end{eqnarray}
where in the last line we have substituted $-m\mu^{1-\eb}$ for
$\lambda$.  The presence of the $1/\ep^2$ term indicates, as expected,
that the leading log term is $\log^2$.  Substituting in $t$ and
$R^{1/2}$, subtracting the infinite terms, and dropping all but the
$\log^2$ terms, we find
\begin{equation}
M(1) = {t^2 \over R^{1/2}} \left( 2\pi^2 \Omega^2 + O(\Omega ) \right).
\end{equation}
Comparing with \ref{scale} we find the coefficients of the leading log's
at second order agree.

The ability to reproduce the non-trivial behavior of the Ising spin
correlators provides us with a degree of confidence that our expression
for the correlators away from $\bh =1 $ are correct.

\subsection{Gross-Neveu SU(2)}
\setcounter{equation}{0}
We now go on to demonstrate that our methods of evaluating the
sine-Gordon correlators reproduce the known behaviour of the
Gross-Neveu model
(sine-Gordon at $\bhs = 2$).  Specifically we demonstrate that we are able
to reproduce the known $\beta$-functions for $\lambda$ and $\hat\beta$
governing the Kosterlitz-Thouless transition to lowest order,
as say calculated
by Amit et al. \cite{amit} and Boyanovsky \cite{boy}.

As shown in section 3, the sine-Gordon theory develops UV singularities as
$\bhs \rightarrow 2$.  As with Ising at $\hat\beta = 1$, these singularities
manifest themselves as poles in gamma functions.  And similiarly to Ising, we
regulate the correlators by continuing away from $\bhs=2$ via
$\bhs = 2 - \epsilon$.  However unlike Ising, we handle these singularites,
because they are UV, with a conventional renormalisation.

In order to facilitate comparison with Amit et al. \cite{amit} and Boyanovsky
\cite{boy} we employ a sine-Gordon action equivalent to theirs:
\begin{equation}
S = -{1 \over 4\pi} \int d^2z \left[ \partial_z \Phi \partial_{\bar z} \Phi
+ {2\lambda \over \bhs} cos(\betah \Phi ) \right] ,
\end{equation}
where the free propagator is
\begin{equation}
\langle \Phi (z) \Phi (0) \rangle = - \log (4 z {\bar z} ) .
\end{equation}
The difference in the propagators amounts to a difference in how its IR
singularities are cured.

To renormalise the theory we
employ the following renormalisation prescription
\begin{eqnarray}
\nn \Phi_0^2 & = & Z_\Phi \Phi_R^2 ; \\
\hat\beta^2_0 & = & Z^{-1}_\Phi \hat\beta^2_R ; \\
\nn ( \cos (\hat\beta\Phi ) )_0 & = & Z_C (\cos (\hat\beta \Phi ) )_R ; \\
\nn \lambda_0 & = & Z_C^{-1} Z_\lambda Z_\Phi^{-1} \lambda_R ;
\end{eqnarray}
where the bare quantities
are listed on the left hand side and the renormalized
quantities on the right hand side.  $\lambda_R$ is understood to be
dimensionless.  This prescription, like Amit et al.'s and
Boyanovsky's, involves a trivial renormalisation of the $\betah$
so as to ensure
$\hat\beta_0 \Phi_0 = \hat\beta_R \Phi_R$.  However
it differs from their's in that it
introduces a wavefunction renormalisation of the $\cos (\hat\beta \Phi )$.
That this
renormalisation is necessary is evident from considering the correlator
$\langle \cos (\hat\beta \Phi (z) ) \cos (\hat\beta \Phi (0) ) \rangle$.  It
possesses singularities at $O(\lambda^2)$, the only way
which to remove is through a
wavefunction renormalisation.
A consequence of this is that the $\beta$-function
for $\lambda$ has been computed incorrectly by both these authors.  (We note
that Amit et al.'s \cite{amit} calculation of
$\beta_\lambda$ at higher orders was already
known to be flawed because of his method of regulating the theory.
See C. Lovelace \cite{love}
for a detailed discussion.)  However as we are interested in
reproducing the $\beta$-functions only at lowest order as a check on our
methodology, we will not concern ourselves here with correcting this error.

The wavefunction renormalisation for $\Phi$, $Z_\Phi$, is found through
calculating the correlator
\begin{equation}
\nn
\langle \partial_z \Phi (z) \partial_z \Phi (0) \rangle .
\end{equation}
Because this correlator is even in $\lambda$, $Z_\Phi$ takes the form
$Z_\Phi = \sum^{\infty}_{n=0} a_{2n} \lambda^{2n}$.
Wavefunction
renormalisation of $\cos (\hat\beta \Phi )$ may be computed through the
correlator $\langle \cos (\hat\beta \Phi (z) ) \cos (\hat\beta \Phi (0) )
\rangle$.
Knowing the wavefunction renormalisations of $\cos (\hat \beta \Phi)$
and of $\Phi$ allows the computation of the coupling constant renormalisation,
$Z_\lambda$, through the correlator $\langle \partial_z \Phi
\cos (\hat\beta\Phi )
\rangle$.  Because $\langle \partial_z \Phi \cos (\hat\beta\Phi )
\rangle$ only
has terms odd in $\lambda$ (and is finite at $O(\lambda )$ ), $Z_\lambda$
takes the form
\begin{equation}
Z_\lambda = \mu^{-\epsilon} ( 1 + \sum^{\infty}_{n=1} a_{2n} \lambda^{2n} ) .
\end{equation}
$\mu$ is an arbitrary mass parameter introduced to make the renormalised
coupling dimensionless.  Here we will not be interested in
computing the higher
order terms.  So we have (rather trivially)
\begin{equation}
Z_\lambda = \mu^{-\epsilon} ( 1+ O(\lambda^2) ) ,
\end{equation}
and we are left to compute $Z_\Phi$.  (We do not need to worry over $Z_C$ as
it contributes to the renormalisation of $\lambda$ at $O(\lambda^2)$, and so
makes a contribution to $\lambda$'s $\beta$-function only at $O(\lambda^3)$.)

To compute $Z_\Phi$, we employ a trick to calculate
$\langle \partial_z \Phi (z)
\partial_{z} \Phi (0) \rangle$.  We can write
\begin{equation}
\langle \partial_z \Phi (z) \partial_{z} \Phi (0) \rangle
= \lim_{\alpha \rightarrow 0} {1 \over \alpha^2} \partial_x \partial_y
\langle e^{i\alpha \Phi (x)} e^{-i\alpha \Phi (y)}
\rangle |_{{x=z}\atop{y=0}} .
\end{equation}
Thus we can use our results from Section 4.  Doing so and taking the $\alpha
\rightarrow 0$ limit we find
\begin{eqnarray}
\langle \partial_z \Phi_0 (z) \partial_{z} \Phi_0 (0) \rangle & = &
-(z)^{-2} - \\
\nn && \hspace{.25in}
{\lambda_0^2 \over 16 \cdot 2^{2\hat\beta_R^2}\pi^2\hat\beta_R^4}
(2-\hat\beta_R^2)(1-\hat\beta_R^2)
(z)^{-\hat\beta_R^2}(\bar z)^{2-\hat\beta_R^2} L ,
\end{eqnarray}
where $L$ is
\begin{equation}
L = - 8\pi\hat\beta_R^2 s(\hat\beta_R^2 ) (\hat\beta_R^2 - 1)
{\Gamma^2 (1-\hat\beta_R^2 ) \Gamma (\hat\beta_R^2 -1)
\Gamma (2-\hat\beta_R^2 )
\over \Gamma^2 (3-\hat\beta_R^2 )}.
\end{equation}
Writing $\hat\beta_R^2 = 2 - \epsilon$,
the singularities in the correlator become
\begin{equation}
\langle \partial_z \Phi (z) \partial_{z} \Phi (0) \rangle
= -(z)^{-2} + {\lambda^2 \over 64\epsilon} (z)^{-\hat\beta_R^2}
(\bar z)^{2-\hat\beta_R^2} \left( 1 + \epsilon (1/2 + 2\log (2) ) \right) .
\end{equation}
Using minimal subtraction, we find $Z_\Phi$ to be
\begin{equation}
Z_\Phi = 1 - {\mu^{-2\epsilon}\lambda_0^2 \over 64 \epsilon} + O(\lambda^4) .
\end{equation}
Knowing $Z_\Phi$ and $Z_\lambda$ allows us to calculate the $\beta$-functions
for $\beta$ and $\lambda$:
\begin{eqnarray}
\beta_\lambda & = & \mu {\partial \lambda_R \over \partial \mu} =
\mu \lambda_0
{\partial \mu^{-\epsilon} \over \partial \mu} = -\epsilon \lambda_R +
O(\lambda^3) ;\\
\beta_{\hat\beta} & = & \mu {\partial \hat\beta_R \over \partial \mu} =
\mu \hat\beta_0 {\partial Z^{1/2}_\Phi \over \partial \mu} =
{\hat\beta_R \lambda_R^2 \over 64}
+ O(\lambda^4) .
\end{eqnarray}
Defining $\delta = - \epsilon/2$, we can recast the above in a form directly
comparable with Amit et al.'s \cite{amit} and Boyanovsky's \cite{boy} result
\begin{eqnarray}
\beta_\lambda & = & 2\delta\lambda_R + O(\lambda^3) ; \\
\beta_\delta & = & \hat\beta_R \beta_{\hat\beta}  =
{\hat\beta_R^2 \lambda_R^2 \over 64}
+ O(\lambda^4) = {\lambda_R^2 \over 32} + {\delta\lambda_R^2\over 32}
+ O(\lambda^4 ).
\end{eqnarray}
The leading terms of these $\beta$-functions agree
with their previous results.
Thus we again have shown that this methodology is consistent with other
methods.

\section{Discussion}

We have shown how the short distance expansion of some sine-Gordon
correlation functions is well defined and reasonably tractable.
The ultimate goal is to find some non-perturbative characterization
of these correlation functions.  We make two remarks.  First of all,
unlike the large distance expansion, the short distance expansions
at and away from the free fermion point are of comparable complexity.
Secondly, we have in no way utilized the integrability structure
of the sine-Gordon theory, i.e. factorizable S-matrix, quantum inverse
scattering method, etc. in developing the short distance expansion.
It would be very interesting to understand the consequences of integrability
in this context.

The results presented herein leave two concrete
avenues open for further exploration.
The first arises from our knowledge of the differential equation
the correlators of the vertex operators satisfy at the free fermion point.
Knowing the first set of terms in the perturbative expansion should allow,
through the differential equation, the generation of higher order terms in
the expansion.

The second is a correct evaluation of the $\beta$-function at
$O(\lambda^3)$ for the coupling
$\lambda$ in the SU(2) Gross-Neveu model.  To make this evaluation two
calculations would be necessary.  The wavefunction renormalisation of
$\cos (\betah \Phi )$ would need to be calculated through the correlator
$\langle \cos (\betah\Phi (z) ) \cos (\betah\Phi (0) ) \rangle$.
Knowing this
then allows the calculation of $Z_\lambda$,
a piece of the coupling constant renormalisation, at $O(\lambda^2)$
through the correlator $\langle \partial_z\Phi (z)
\cos (\betah\Phi (0) ) \rangle$.  From there, $\beta_\lambda$ is directly
calculable.

\section{Acknowledgements}
The authors would like to thank P. Lepage for useful discussions.

\pagebreak
\setcounter{equation}{0}
\appendix
\section{A Generalized Euler Integral}
In this appendix we evaluate the following generalized Euler integral,
\begin{eqnarray}
I & = & \int^1_0 dw_1 w_1^{a_1} (1-w_1)^{b_1} \times \\
& & \nn \int^{w_1}_0 dw_2 w_2^{a_2} (1-w_2)^{b_2} (w_1 - w_2)^{\alpha_{12}}
\times \\
& & \nn \int^{w_2}_0 dw_3 w_3^{a_3} (1-w_3)^{b_3} (w_1 - w_3)^{\alpha_{13}}
(w_2 - w_3)^{\alpha_{23}} \times \\
& & \nn \hspace{2in} \vdots \\
& & \nn \int^{w_{n-1}}_0 dw_n w_n^{a_n} (1-w_n)^{b_n}
(w_1 - w_n)^{\alpha_{1n}}
\cdots (w_{n-1} - w_n)^{\alpha_{n-1,n}}.
\end{eqnarray}

\noindent Employing the following series of changes of variables,
$$ w_2 = w_2'w_1,~ w_3 = w_3'w_2'w_1 ,~ \ldots ~, ~w_n = w_n'\cdots w_2'
w_1 ,$$
the above integral reduces to
\begin{eqnarray}
I & = & \int^1_0 dw_1 w_1^{\sum^n_{i=1} a_i + \sum^n_{i < j}
\alpha_{ij}+n-1}
(1-w_1)^{b_1} \times \\
\nn & & \int^1_0 dw_2 w_2^{\sum^n_{i=2} a_i + \sum^n_{1 < i < j}
\alpha_{ij} +n-2}
(1-w_1w_2)^{b_2} (1-w_2)^{\alpha_{12}} \times \\
\nn & & \int^1_0 dw_3 w_3^{\sum^n_{i=3} a_i + \sum^n_{2 < i < j}
\alpha_{ij} +n-3}
(1-w_1w_2w_3)^{b_3} (1-w_2w_3)^{\alpha_{13}} \times \\
\nn & & \hspace{3.4in} (1-w_3)^{\alpha_{23}} \times \\
\nn & & \hspace{2in} \vdots \\
\nn & & \int^1_0 dw_n w_n^{a_n}
(1-w_1\cdots w_n)^{b_n} (1-w_2\cdots w_n)^{\alpha_{1n}}
(1-w_3\cdots w_n)^{\alpha_{2n}} \cdots \\
\nn & & \hspace{2.5in} (1-w_{n-1}w_n)^{\alpha_{n-2,n}}
(1-w_n)^{\alpha_{n-1,n}}.
\end{eqnarray}
\noindent Using the expansion,
\begin{equation}
(1-w)^a = \sum^\infty_{k=0} {(-a)_k \over k!} w^k ,
\end{equation}
where $(x)_k \equiv x(x+1) \cdots (x+k-1) \equiv \Gamma(x+k)/\Gamma(x)$,
the above becomes
\vfill\eject
\begin{eqnarray}
I & = & \int^1_0 dw_1 w_1^{\sum^n_{i=1} a_i +
\sum^n_{i < j} \alpha_{ij}+n-1} (1-w_1)^{b_1} \times \\
\nn & & \sum_{k_{12}} {(-b_{2})_{k_{12}} \over k_{12}!}
\int^1_0 dw_2 w_2^{\sum^n_{i=2} a_i + \sum^n_{1 < i < j}\alpha_{ij}+n-2}
(1-w_2)^{\alpha_{12}} (w_1w_2)^{k_{12}} \times \\
\nn & & \sum_{k_{23} \atop k_{123}}
{(-b_3)_{k_{123}}(-\alpha_{13})_{k_{23}} \over k_{123}! k_{23}!}
\int^1_0 dw_3 w_3^{\sum^n_{i=3} a_i + \sum^n_{2 < i < j}\alpha_{ij} +n-3}
(1-w_3)^{\alpha_{23}} \times \\
\nn & & \hspace{2.5 in} (w_1w_2w_3)^{k_{123}} (w_2w_3)^{k_{23}} \times \\
\nn & & \hspace{2in} \vdots \\
\nn & & \sum_{{k_{1\cdots n} \atop {k_{2\cdots n} \atop
{\vdots \atop k_{n,n-1}}}}}
{(-b_n)_{k_{1\cdots n}} (-\alpha_{1n})_{k_{2\cdots n}}
(-\alpha_{2n})_{k_{3\cdots n}} \cdots (-\alpha_{n-2,n})_{k_{n-1,n}}
\over k_{1\cdots n}! k_{2\cdots n}! \cdots k_{n,n-1}! } \\
\nn & & \hspace{.5in}\times\int^1_0 dw_n w_n^{a_n} (1-w_n)^{\alpha_{n,n-1}}
(w_1\cdots w_n)^{k_{1\cdots n}} \cdots (w_nw_{n-1})^{k_{n,n-1}}.
\end{eqnarray}
where the $k_{m \cdots n}$ arise from the expansion of $(1-w_m\cdots w_n)$.
\noindent Defining the following notation
\begin{eqnarray}
\sigma_m  & = & \sum_{i=m}^n a_i + \sum^n_{i \neq j > m -1} \alpha_{ij}
+ (n - m) , \\
\sum_m k  & = & {\rm sum~of~k's~with~index~m} \\
\nn & = & k_{1\cdots m} + k_{1\cdots m,m+1} + \cdots +
k_{1\cdots m \cdots n} + \\
\nn & & k_{2\cdots m} + k_{2\cdots m,m+1} + \cdots +
k_{2\cdots m \cdots n} +\\
\nn & & \hspace{1in} \vdots \\
\nn & & k_{m,m+1} + k_{m,m+1,m+2} + \cdots + k_{m \cdots n},
\end{eqnarray}
and
\begin{eqnarray}
(k)! & = & k_{12}! k_{123}! \cdots k_{1\cdots n}! \times
k_{23}! \cdots k_{2\cdots n}! \times \cdots \times k_{n,n-1}! ,
\end{eqnarray}
\noindent I simplifies to
\vfill\eject
\begin{eqnarray}
I & = & {\sum_{\{k\}} \prod^n_{i=2} (-b_i)_{k_{1\cdots i}}
\prod^n_{i < j-1} (-\alpha_{ij})_{k_{i+1\cdots j}} / (k)!} \times \\
\nn & & \int^1_0 dw_1 w_1^{\sigma_1 + \sum_1 k} (1-w_1)^{b_1} \times \\
\nn & & \int^1_0 dw_2 w_2^{\sigma_2 + \sum_2 k} (1-w_1)^{\alpha_{12}} \times
\\
\nn & & \hspace{1.5in} \vdots \\
\nn & & \int^1_0 dw_n w_n^{\sigma_n + \sum_n k} (1-w_1)^{\alpha_{n-1,n}}
\end{eqnarray}
Doing these integrals, we are able to put I into its final form:
\begin{eqnarray}
\nn I & = & \B(b_1+1,\sigma_1+1)\B(\alpha_{12}+1,\sigma_2+1)\cdots
B(\alpha_{n-1,n}+1,\sigma_n+1) \times \\
\nn & & \hspace{.5in} {\sum_{\{k\}} \prod^n_{i=2} (-b_i)_{k_{1\cdots i}}
\prod^n_{i < j-1} (-\alpha_{ij})_{k_{i+1\cdots j}} / (k)!} \times \\
& & \hspace{1.0in} {(\sigma_1+1)_{\sum_1 k} / (\sigma_1 + b_1 +
2)_{\sum_1 k}} \times \\
\nn & & \hspace{1.0in} {(\sigma_2+1)_{\sum_2 k} / (\sigma_1 + \alpha_{12} +
2)_{\sum_2 k}} \times \\
\nn & & \hspace{1.5in} \vdots \\
\nn & & \hspace{1.0in} {(\sigma_n + 1)_{\sum_n k} / (\sigma_n + \alpha_{n-1,n}
+ 2)_{\sum_2 k}}.
\end{eqnarray}

\section{Evaluation of K(i,j)}
\setcounter{equation}{0}

In this appendix we evaluate the integrals $K(i,j)$ introduced in Section
4.2.  Using Appendix A we can write $K(i,j)$ as
\begin{eqnarray}
\nn K(i,j) & = & \int^1_0 dw_i w_i^{-2 + \hat\beta^2}(1-w_i)^{\gamma_i}
\int^{w_i}_0 dw_j w_j^{-2 + \hat\beta^2}(1-w_j)^{\gamma_j}
(w_i - w_j)^{-\hat\beta^2}\\
& = & \B (\gamma_i +1,\bhs -2) \B(1-\bhs , \bhs -1) \times \\
\nn & & \sum_{k=0}^{\infty} {(-\gamma_j)_k (\bhs -2)_k (\bhs -1)_k \over k!
(\gamma_i + \bhs  -1)_k (0)_k}
\end{eqnarray}
where $\B(x,y) = \Gamma (x) \Gamma (y) / \Gamma(x+y)$, $(a)_k = \Gamma
(a+k)/\Gamma(a)$, and the sum $\sum_k$ is the standard
hypergeometric function,
\begin{equation}
\gh (-\gamma_j,\bhs - 2,\bhs - 1,\gamma_i + \bhs - 1,0, 1) .
\end{equation}
However this form is somewhat problematic as
a pole in \gh ~is multiplying a zero in \B.  To make
sense of this expression we continue $w_j^{-2+\hat\beta^2}$ to
$w_j^{-2+\hat\beta^2 +\epsilon}$ in the above integrand and take
the limit $\epsilon \rightarrow 0$.  With this continuation,
$K(i,j)$ equals
\begin{eqnarray}
K(i,j) = \B (\gamma_i +1,\bhs -2) \B(1-\bhs , \bhs -1 + \epsilon) \times \\
\gh (-\gamma_j,\bhs - 2,\bhs - 1,\gamma_i + \bhs - 1,\epsilon, 1) .
\end{eqnarray}
Using the identity
\begin{eqnarray}
\lefteqn{\lim_{\zeta \rightarrow -n}
{\gh(\alpha,\beta,\gamma,\delta,\zeta,z) \over \Gamma(\zeta)} = } \\
& & \nn {\alpha (\alpha + 1) \cdots (\alpha + n) \beta (\beta + 1) \cdots
(\beta + n) \gamma (\gamma + 1) \cdots (\gamma + n) \over (n+1)!
\delta (\delta + 1) \cdots (\delta +n)} \times \\
& &  \nn z^{n+1}
\gh (\alpha +n+1,\beta +n+1,\gamma +n+1,\delta +n+1,2,z) ,
\end{eqnarray}
and then taking the limit $\epsilon \rightarrow 0$, we obtain
\begin{eqnarray}
\nn \lefteqn{K(i,j) = B(1+\gamma_i,\bhs-1)
\Gamma(1-\bhs)\Gamma(\bhs-1)(-\gamma_j)(\bhs -1) \times} \\
& & \hspace{1.2in} \gh(1-\gamma_j,\bhs -1,\bhs ,\bhs +\gamma_i, 2, 1),
\end{eqnarray}
\noindent where we have also used the relation $x\Gamma (x) = \Gamma (x+1)$.

\section{Analytic Continuation of \gh}
\setcounter{equation}{0}

The Gaussian hypergeometric function $\gh(\alpha ,\beta , \gamma , \delta
, \zeta ,1)$ is, in general, a meromorphic
function of the variables $\alpha ,\beta ,\gamma , \delta$, and $\zeta$.
However its sum representation (as say given in Appendix B)
is not necessarily convergent for all values of these variables.
The convergence of the sum is determined by defining the ``excess''
parameter, s, by
\begin{equation}
s = \delta + \zeta - (\alpha + \beta + \gamma).
\end{equation}
If $s>0$ the sum converges.  However we will often encounter situations
where $s < 0$ and the sum representation is no good.  To get around this
problem analytic continutation formulae are needed.  The only one that will
prove to be necessary is an identity known as Dixon's theorem:

\begin{equation}
\gh (\alpha ,\beta ,\gamma , \delta ,\zeta ,1) =
{\Gamma (\delta ) \Gamma (\zeta ) \Gamma (s) \over \Gamma (\alpha)
\Gamma (s + \beta) \Gamma (s + \gamma)}
\gh (\delta - \alpha ,\zeta - \alpha ,s,s+\beta ,s+\gamma ,1).
\end{equation}

\noindent The convergence
of the hypergeometric function of the right hand side is
then determined by the positivity of $\alpha$.  Notice that because \gh
{}~is invariant under permutations of $\alpha$, $\beta$, and $\gamma$
and permutations of $\delta$ and $\zeta$, variations (144 in total - not
all necessarily distinct) of
this identity exist.

\section{Taking $\bh \rightarrow 1$ in
$\langle \psi_-(z,\bar z)\psi_+(0) \rangle$}
\setcounter{equation}{0}
We focus on the Taylor series expansions of the hypergeometric
functions in $K_2(\pm\bh )$.  The expansion of the
$\gh$ function in $K_2(\bh )$ is given by
\begin{eqnarray}
\nn\gh (1+\bhs ,\bhs -1 ,2-\bhs ,2,1,1) & = &
{\Gamma (1-\bhs ) (1-\bhs)^2(-\bhs ) \over \Gamma (1+\bhs ) \Gamma (3 - 2\bhs)
(3-2\bhs )} \times \\
&& \hspace{-2.0in} \gh (2-\bhs ,1-\bhs ,2-\bhs ,2,4-2\bhs ,1) ,
\end{eqnarray}
where we have used Dixon's theorem, C.2, and the identity, B.2.  Evaluating
further we have
\begin{eqnarray}
\nn\gh (1+\bhs ,\bhs -1 ,2-\bhs ,2,1,1) & = &
{\Gamma (2-\bhs ) (1-\bhs )(-\bhs ) \over \Gamma (1+\bhs ) \Gamma (3 - 2\bhs)
(3-2\bhs )} \times \\
&& \hspace{-2in} \gh (2-\bhs ,1-\bhs ,2-\bhs ,2,4-2\bhs ,1) \\
\nn & = & \epsilon + O (\epsilon^2 ).
\end{eqnarray}
The $\gh$ function in $K_2(-\bh )$ is simpler to evaluate:
\begin{eqnarray}
\nn \gh (1-\bhs , \bhs -1,2-\bhs ,2,1,1) & = &
\gh (-\ep ,\ep ,1+\ep,2,1,1) \\
& = & 1 + O(\ep^2 ).
\end{eqnarray}

\section{Taking $\bh \rightarrow 1$ in
$\langle \sigma (z,\bar z)\sigma (0) \rangle^2$}
\setcounter{equation}{0}

As in appendix D, we focus only on the hypergeometric functions
that appear in the expressions that are being expanded about $\bh
= 1$.  Appearing in $K_{\pm}(\bh )$ we have the function:
\begin{equation}
\sk = \gh (1\pm\bhs /2,\bhs ,\bhs -1,\bhs \pm \bhs /2,2,1).
\end{equation}
To $O(\ep )$, $\sk$ is
\begin{eqnarray}
\nn \sk & = & \gh(1\pm 1/2,1,\ep ,1\pm 1/2,2,1) + O(\ep^2) \\
\nn & = & \sum^{\infty}_{k=0} {(\ep )_k (1)_k \over (2)_k k!} + O(\ep^2 )
\\
\nn & = & {\Gamma (2) \Gamma (1-\ep ) \over \Gamma (2-\ep ) \Gamma (1)}
+O(\ep^2) \\
& = & 1 + \ep + O(\ep^2) ,
\end{eqnarray}
where we have used the Gaussian summation formula
\begin{equation}
_2F_1(a,b,c,1) =
\sum^{\infty}_{k=0} {(a)_k (b)_k \over (c)_k k!} = {\Gamma (c)
\Gamma (c-a-b) \over \Gamma (c-a) \Gamma (c-b)}.
\end{equation}

In $J_{\pm}$ we face the sum
\begin{equation}
\sj = \gh (\pm \bhs /2,2-\bhs ,1\pm\bhs /2,3-\bhs\pm\bhs /2,2-\bhs\pm\bhs
/2,1).
\end{equation}
Using the analytic continuation formula in appendix C, $\sj$ can be
transformed into
\begin{equation}
\sj = {\Gamma (3-\bhs \pm \bhs /2) \Gamma (2-\bhs ) \over
\Gamma (1\pm\bhs /2) \Gamma (4-2\bhs )} \nsj
\end{equation}
where $\nsj$ is
\begin{equation}
\nsj = \gh (1-\bhs ,2-\bhs ,2-\bhs ,2-\bhs \pm\bhs /2,4-2\bhs ,1).
\end{equation}
To $O(\ep )$, $\nsj$ is
\begin{eqnarray} \label{three}
\bar S_{J_{\pm}}(\ep) & = & \gh (-\ep ,1,1,1\pm 1/2,2,1) + O(\ep^2)\\
\nn & = & 1 - \ep\sum^{\infty}_{k=1}{(1)_{k-1}(1)_k (1)_k \over
k! (1\pm 1/2)_k (2)_k} .
\end{eqnarray}
The evaluations of the two sums in \ref{three} are simlar.  So we will
only be explicit in the evaluation of $\spl$.

We may write
\begin{eqnarray}
\spl & = & 1 - \ep \sum^{\infty}_{k=1} {(1)_{k-1} (1)_k \over (3/2)_k
(2)_k }\\
\nn & = & 1 - \ep \lim_{\delta \rightarrow 0} \sum^{\infty}_{k=2}
{(1+\delta )_{k-2} (1+\delta )_{k-1} \over (3/2)_{k-1} (2)_{k-1} }\\
\nn & = & 1 -{\ep \over 2} \lim_{\delta \rightarrow 0} {\Gamma (\delta -1)
\Gamma (\delta ) \over \Gamma^2(1 +\delta ) }
\sum^{\infty}_{k=2} {(\delta - 1)_k (\delta )_k \over
(1/2)_k k!}.
\end{eqnarray}
Now using the Gaussian summation formula we find
\begin{equation}
\spl = 1 - {\ep \over 2} \lim_{\delta \rightarrow 0 }
{\Gamma (\delta -1) \Gamma (\delta ) \over \Gamma^2(1+\delta )}
\left[ {\Gamma (3/2 -2\delta ) \Gamma (1/2) \over \Gamma (3/2-\delta )
\Gamma (1/2 -\delta ) } - 1 - 2\delta (\delta - 1) \right]
\end{equation}
Evaluating this limit we obtain
\begin{equation}
\spl = 1 + \ep ({\pi^2 \over 4} - 3) + O(\ep^2 ).
\end{equation}
Similarly we find for $\sm$
\begin{equation}
\sm = 1 - \ep ({\pi^2 \over 2} + 2) + O(\ep^2 ).
\end{equation}
\pagebreak


\begin{thebibliography}{99}
\bibitem{cpt}A. B. Zamolodchikov, Int. J. Mod. Phys. A4 (1989) 4235.
\bibitem{smir}F. A. Smirnov, {\em Form Factors in Completely Integrable
Models of Quantum Field Theory}, in Advanced Series in Mathematical
Physics 14, World Scientific, 1992.
\bibitem{korepin}V. E. Korepin, N. M. Bogoliubov and A. G. Izergin,
{\em Quantum Inverse Scattering Method and Correlation Functions},
Cambridge University Press, Cambridge, 1993.
\bibitem{iz}J. Zuber and C. Itzykson, Phys. Rev. D 15 (1977) 2875.
\bibitem{wu}T.T. Wu, B.M. McCoy, C.A. Tracy, and E. Barouch, Phys. Rev.
B13 (1976) 316.
\bibitem{bb} O. Babelon and D. Bernard, Phys. Lett. B288 (1992) 113.
\bibitem{ble} D. Bernard and A. LeClair, {\em Differential Equations for
Sine-Gordon Correlation Functions at the Free Fermion Point},
CLNS 94/1276, SphT-94-021, hep-th/9402144, to appear in Nucl. Phys. B.
\bibitem{cole} S. Coleman, Phys. Rev. D 11 (1975) 2088.
\bibitem{dot} Vl. S. Dotsenko, Nucl. Phys. B 240 (1989) 687.
\bibitem{mand} S. Mandelstan, Phys. Rev. D 11 (1975) 3026.
\bibitem{Zam} A. B. Zamolodchikov, {\em Mass Scale in Sine-Gordon and its
Reductions}, LPM-93-06.
\bibitem{amit} D. J. Amit, Y. Y. Goldschmidt, and G. Grinstein, J. Phys. A
13 (1980) 585.
\bibitem{boy}  D. Boyanovsky, J. Phys. A 22 (1989) 2601.
\bibitem{love}  C. Lovelace, Nucl. Phys. B 273 (1986) 413.
\end{thebibliography}
\end{document}